\documentclass[sn-basic,iicol]{sn-jnl}
\pdfoutput=1
\usepackage[english]{babel}
\usepackage[utf8]{inputenc}
\usepackage[T1]{fontenc}
\usepackage{natbib}
\setcitestyle{comma,square,numbers}

\usepackage{graphicx}
\usepackage{tikz}
\usepackage{psfrag}
\usepackage{color}
\usepackage{xcolor}

\graphicspath{{./figures/}}
\usepackage[hypcap=true, font=footnotesize]{caption}
\usepackage[hypcap=true]{subcaption}
\usepackage{float}
\usepackage{listliketab}
\usepackage{tabularx}
\usepackage{array}
\usepackage{longtable}
\usepackage{booktabs}
\usepackage{multirow}
\usepackage{paralist}
\newcolumntype{Y}{>{\raggedright\arraybackslash}X}
\newcolumntype{L}[1]{>{\raggedright\let\newline\\\arraybackslash\hspace{0pt}}p{#1}}

\usepackage{amsmath}
\usepackage{amssymb}
\usepackage{amsfonts}
\usepackage{amsxtra}
\usepackage{mathtools}
\usepackage{empheq}
\usepackage{stmaryrd}
\usepackage{icomma}
\usepackage{relsize}

\usepackage{hyperref}
\hypersetup{hidelinks}

\newcommand{\dt}{\,\mathrm{d}\tau}
\newcommand{\dn}[1]{\overset{\!\!\!\!\!\raisebox{-1pt}{~\tikz\draw[line width=0.4pt]circle(0.7pt);~}\!\!\!\!\!}{#1}}
\newcommand{\ddn}[1]{\overset{\!\!\!\!\!\raisebox{-1pt}{~\tikz\draw[line width=0.4pt]circle(0.7pt);~}\!\!\!\!\raisebox{-1pt}{~\tikz\draw[line width=0.4pt]circle(0.7pt);~}\!\!\!\!\!}{#1}}
\newcommand{\grad}{\boldsymbol{\nabla}}

\newcommand{\norm}[2]{\|#1\|_{#2}}

\newcommand{\bs}[1]{\boldsymbol{#1}}

\makeatletter
\newcommand*\symboldummy{\mathpalette\symboldummy@{.625}}
\newcommand*\symboldummy@[2]{\mathbin{\vcenter{\hbox{\scalebox{#2}{$\m@th#1\bullet$}}}}}
\makeatother

\begin{document}

\title{An adaptive acceleration scheme for phase-field fatigue computations} 

\author[1]{\fnm{Jonas} \sur{Heinzmann}}\email{jheinzmann@ethz.ch}
\author[1]{\fnm{Pietro} \sur{Carrara}}\email{pcarrara@ethz.ch}
\author[2]{\fnm{Marreddy} \sur{Ambati}}\email{marreddy.ambati@ge.com}
\author[3]{\fnm{Amir Mohammad} \sur{Mirzaei}}\email{amir.mirzaei@polito.it}
\author*[1]{\fnm{Laura} \spfx{De} \sur{Lorenzis}}\email{ldelorenzis@ethz.ch}

\affil*[1]{\orgdiv{Department of Mechanical and Process Engineering}, \orgname{ETH Z\"urich}, \orgaddress{\street{Tannenstrasse 3}, \city{Z\"urich}, \postcode{8092}, \country{Switzerland}}}
\affil[2]{\orgdiv{Material Mechanics and Durability}, \orgname{GE Global Research}, \orgaddress{\street{1 Research Circle}, \city{Niskayuna}, \postcode{12309}, \state{New York}, \country{USA}}}
\affil[3]{\orgdiv{Department of Structural, Geotechnical and Building Engineering}, \orgname{Politecnico di Torino}, \orgaddress{\street{Corso Duca degli Abruzzi 24}, \city{Torino}, \postcode{10129}, \country{Italy}}}

\subsection*{Graphical abstract}
\begin{center}
    \includegraphics[width=\textwidth]{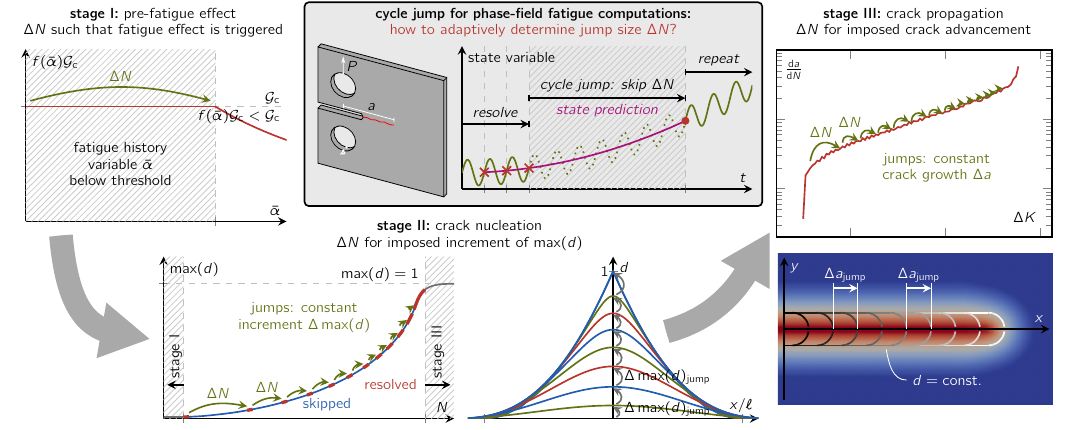}
\end{center}

\abstract{
    Phase-field models of fatigue are capable of reproducing the main phenomenology of fatigue behavior.
    However, phase-field computations in the high-cycle fatigue regime are prohibitively expensive due to the need to resolve \textit{spatially} the small length scale inherent to phase-field models and \textit{temporally} the loading history for several millions of cycles.
    As a remedy, we propose a fully adaptive acceleration scheme  based on the cycle jump technique, where the cycle-by-cycle resolution of an appropriately determined number of cycles is skipped while predicting the local system evolution during the jump.
    The novelty of our approach is a cycle-jump criterion to determine the appropriate cycle-jump size based on a target increment of a global variable which monitors the advancement of fatigue.
    We propose the definition and meaning of this variable for three general stages of the fatigue life.
    In comparison to existing acceleration techniques, our approach needs no parameters and bounds for the cycle-jump size, and it works independently of the material, specimen or loading conditions. Since one of the monitoring variables is the fatigue crack length, we introduce an accurate, flexible 
 and efficient method for its computation, which overcomes the issues of conventional crack tip tracking algorithms and enables the consideration of several cracks evolving at the same time.
    The performance of the proposed acceleration scheme is demonstrated with representative numerical examples, which show a speedup reaching up to four orders of magnitude in the high-cycle fatigue regime with consistently high accuracy.
}


\maketitle

\renewcommand{\thefootnote}{\textcolor{white}{\arabic{footnote}}} 
\footnotetext{Preprint submitted to Computational Mechanics\\(published version: \href{https://doi.org/10.1007/s00466-024-02551-8}{doi.org/10.1007/s00466-024-02551-8})}
\renewcommand{\thefootnote}{\textcolor{black}{\arabic{footnote}}}

\section{Introduction}\label{sec:introduction}
Predictive modeling of fatigue fracture is of interest for a wide range of applications as fatigue is the most common cause of failure for many engineering components \cite{hosseini_theoretical_2018}.
The phase-field modeling approach, originally proposed for brittle fracture under monotonically increasing loads \cite{bourdin_numerical_2000}, was recently extended to fatigue (see \cite{alessi_phenomenological_2018,carrara_framework_2020,seiler_efficient_2020,schreiber_phase_2020,amendola_thermomechanics_2016,grossman-ponemon_phase-field_2022} among many others) and proved able to reproduce the main phenomenological features of fatigue behavior, namely the Wöhler curve, describing the cycle count until failure as a function of the mean load, and the crack growth rate curve, giving the crack propagation rate versus the stress intensity factor amplitude.
Phase-field models of fracture are endowed with an internal length scale which is significantly smaller than the characteristic dimensions of the domain under study. In the discretized setting, using e.g. the finite element method, the need to resolve this length leads to fine meshes and thus high computational cost already in quasi-static fracture computations. 
While various techniques can be used to accelerate the solution of individual time steps, e.g. adaptive mesh refinement (e.g. \cite{gupta_adaptive_2022,hennig_projection_2018,freddi_adaptive_2023}), special element formulations (e.g. \cite{olesch_adaptive_2021,ambati_phasefield_2022}) or solution algorithms (e.g. \cite{gerasimov_line_2016,kristensen_phase_2020,chen_fft_2019}),
an additional issue in fatigue computations is the need to resolve possibly millions of loading cycles. Thus, while computations in the low-cycle fatigue (LCF) regime may still be feasible, computations in the  high-cycle fatigue (HCF) and very high-cycle fatigue (VHCF) regimes become impracticable.

\begin{figure}[t]
    \centering
    \includegraphics{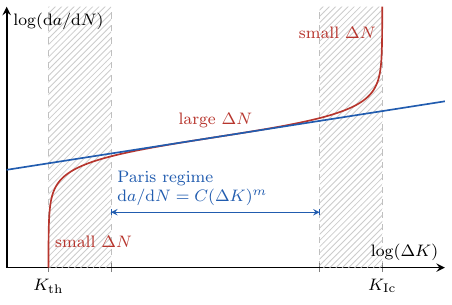}
    \caption{Illustrative crack growth rate curve showing the different regimes of fatigue crack propagation and their relation with a proper cycle jump size $\Delta N$: larger within highly non-linear regimes and smaller during the stable crack propagation or Paris regime.}
    \label{fig:cgr_adaptivity}
\end{figure}

Although the slow system evolution within a cycle can be exploited to accelerate the computation of an individual cycle, as done e.g. in \cite{kristensen_accelerated_2023}, the biggest potential for speedup is in the resolution of the cycle domain.
Next to multi-time-scale approaches, which aim at separating the processes taking place at different temporal scales (evolution within an individual cycle vs. evolution from cycle to cycle) or temporal homogenization techniques, such as the ones presented in \cite{lemaitre_twoscale_1999,oskay_fatigue_2004,bhattacharyya_multi-temporal_2018}, the cycle jump method is a commonly adopted method to speed up fatigue computations due to its straightforward implementation and general applicability.
It was first introduced by \citeauthor{lemaitre_damage_1994} for a damage model with cyclic plasticity in \cite{lemaitre_damage_1994}, where the core idea is to `jump cycles', i.e. to skip the resolution of blocks of cycles.
The evolution of the system during the jumped cycles is predicted by extrapolating selected state variables based on their evolution during a sequence of resolved cycles.
The predicted values are then imposed as a new system state, before the procedure restarts, meaning that again, several cycles are computed to have a basis for the extrapolation of the subsequent cycle jump, and so forth.

In the simplest form of the method, the number of cycles to be jumped, $\Delta N$, is the same for each jump, while no specific criterion is used to decide whether it is appropriate or not to jump at any given point in time.
This strategy is applicable only to  slowly (almost linearly) evolving systems, and is typically not practicable for fatigue computations.
This is illustrated in Fig. \ref{fig:cgr_adaptivity}: during the crack nucleation phase, only a small number of cycles can be jumped due to the strongly non-linear behavior of the system; during stable crack propagation, including the Paris regime, a large number of cycles can be jumped; close to failure, again only a small number of cycles can be jumped.
To overcome such limitations, adaptive approaches that tailor the number of cycles to be skipped to the system evolution have been proposed in the literature.
The explicit approaches aim at defining $\Delta N$ based on the computed states of the system up to the time of the jump, and determine the extension of the jump to keep the error on the system evolution within a given tolerance \cite{Cheng_cyclejump_2022,Nesnas_cycle_2000,peerlings_gradient-enhanced_2000,cojocaru_simple_2006,kiewel_application_2000,Moslemian_2011_accelerated}. This strategy was applied to phase-field fatigue models by \citeauthor{seles_general_2021} in \cite{seles_general_2021} and \citeauthor{haveroth_non-isothermal_2020} in \cite{haveroth_non-isothermal_2020}, the latter reporting a four-fold reduction of computational time.
However, the strategy heavily relies on an allowed error estimate, which the user must set based on geometry, material and loading conditions.
Furthermore, these cycle jump approaches typically determine a feasible $\Delta N$ at each node or Gauss point, meaning that, out of all locally determined cycle-jump sizes, one suitable for the global system response has to be selected, typically the minimum \cite{seles_general_2021}. Papagem et al. \cite{paepegem_cycle_2015} propose to choose the global $\Delta N$ based on a percentile of the cumulative distribution of the local cycle-jump sizes.
In \cite{yan_efficient_2022}, within an adaptive cycle stepping scheme, a cycle increment is obtained based on a stage-wise defined allowed damage increment, again relying on a crucial user choice for the latter.
Another possible strategy involves the adoption of implicit algorithms, where the jump extension is computed accounting also for the state of the system after the jump. For instance, in \cite{loew_accelerating_2020} \citeauthor{loew_accelerating_2020} propose to iteratively change the predicted state of the system for a fixed $\Delta N$ until the trapezoidal integration rule used for the extrapolation is satisfied within a given tolerance. The arising non-linear equations are solved using a Newton-Raphson scheme involving the computation of a full cycle for each iteration.
The authors further propose a method to adaptively determine $\Delta N$, which modulates the cycle-jump size based on the number of iterations that the solver needs to reach convergence.
A similar iterative scheme is combined with adaptive mesh refinement techniques by \citeauthor{Jaccon_2023_adaptive} in \cite{Jaccon_2023_adaptive}.
Regardless of being explicit or implicit, the accuracy and speedup of the presented methods critically depend on user-defined parameters, while no method exists to obtain a good estimate for them \textit{a priori}.

To address the issues of existing acceleration schemes, we propose a new adaptive cycle jump (ACJ) algorithm, which is based on the idea of constraining the growth of a representative global variable during the jumped cycles.
To this end, the fatigue life of a component is divided into three stages: (I) an initial stage before fatigue effects are triggered, (II) crack nucleation, and (III) crack propagation up to failure.
During the first stage, the number of cycles to be skipped is determined so as to jump to the point at which the fatigue effect is triggered for the first time.
In the second stage, $\Delta N$ is computed such that a target increment of the maximum value of the phase field is obtained, thus progressively approaching the first full crack development.
Analogously, in the third stage, $\Delta N$ is determined so as to induce a predefined crack length increment.
We introduce the concept of a trial cycle after a cycle jump as an \textit{a posteriori} check of the cycle-jump criterion, which allows to automatically correct too optimistic cycle jumps.
Further, for the third stage, a novel crack tip tracking algorithm is presented which overcomes the dependency on the spatial discretization and hence the inaccuracy of conventional crack growth monitoring strategies.
Using this, the proposed cycle jump scheme can achieve a speedup of up to more than four orders of magnitude while giving consistently low errors with a stable behavior, all without the need to tune any additional parameters.
As better clarified later, we introduce optional parameters allowing to prioritize accuracy over speedup or vice versa.

This paper is structured as follows.
In Section \ref{sec:phase-field_fatigue_modeling}, we briefly overview the phase-field model for brittle fatigue adopted in this paper.
Section \ref{sec:adaptive_acceleration_based_on_cycle_jump_constraint} introduces the ACJ scheme, for which Section \ref{sec:numerically_efficient_crack_tip_tracking} presents the concept of the smeared crack length as crack tip tracking algorithm.
In Section \ref{sec:numerical_examples}, the behavior and properties of the proposed acceleration scheme are demonstrated, and the obtained accuracy and speedup are compared to those of existing acceleration techniques. Conclusions are drawn in Section \ref{sec:conclusions}.

\section{Phase-field modeling of brittle fatigue }\label{sec:phase-field_fatigue_modeling}
The phase-field approach to brittle fracture \cite{bourdin_numerical_2000} was first derived as the regularization of the formulation in \cite{francfort_revisiting_1998}, which recasts \citeauthor{griffith_phenomena_1921}'s energy criterion \cite{griffith_phenomena_1921} into a variational framework. Later on, it was shown that the same formulation (with a more flexible choice of the involved functions) can be constructed as a special family of gradient damage models \cite{pham_approche_2010,pham_approche_2010-1}.
The formulation departs from the total energy functional for a body occupying domain $\Omega$,
\begin{equation}
    \begin{aligned}
        &\mathcal{E} (\bs{u}, d) =\\
        &\hspace{0.5cm}\hphantom{+} \int_\Omega \left[ g(d) \psi^{\text{el}}_+(\boldsymbol{\varepsilon}(\bs{u})) + \psi^{\text{el}}_-(\boldsymbol{\varepsilon}(\bs{u})) \right] \mathrm{d}V\\
        &\hspace{0.5cm}+ \int_\Omega  \frac{\mathcal{G}_{\text{c}}}{ c_w} \left( \frac{w (d)}{\ell} \! + \ell |{\grad d}|^2 \right) \mathrm{d}V\\
        &\hspace{0.5cm}- \int_{\partial \Omega_{\text{N}}} \!\!\! \bar{\bs{t}} \cdot \bs{u} \, \mathrm{d}A \text{ .}
    \end{aligned}
    \label{eq:energy}
\end{equation}
 Here, $\bs{u}$ is the displacement field, $\bs{\varepsilon}$ is the infinitesimal strain tensor, $d \in [0,1]$ is the phase-field or damage variable, $\ell$ is the regularization length, and $\mathcal{G}_{\text{c}}$ is the critical energy release rate. 
 $\psi^{\text{el}}_+$ and $\psi^{\text{el}}_-$ are respectively the active and the inactive part of the elastic strain energy density, with the active part coupled to the phase field by means of the degradation function $g(d) = (1-d)^2 + g_0$, where the residual stiffness $g_0$ is here set to $10^{-6}$. Coupling the phase-field only to $\psi^{\text{el}}_+$ allows to model contact between crack faces and  asymmetric fracture behavior in tension and compression  \cite{de_lorenzis_nucleation_2022,miehe_phase_2010,amor_regularized_2009,lancioni_variational_2009,freddi_regularized_2010,Vicentini2023}. $w(d)$ is the so-called local dissipation function, for which two common choices are
 $w(d) = d$ (known as AT1 model) or $w(d) = d^2$ (AT2 model), with the normalization factor $c_w$ equal to  $8/3$ or $2$, respectively.
Finally, $\bar{\bs{t}}$ is the traction vector acting on the Neumann boundary of the domain, $\partial \Omega_{\text{N}}$.
Body forces as well as inertia terms are not considered in this work. Considering an irreversibility constraint for the damage variable (i.e. that the damage can only increase), the unilateral local minimization of the energy (\ref{eq:energy}) with respect to displacement and damage fields in the time-discrete setting delivers the coupled system of governing equations whose solution yields the time-discrete evolution of the two fields. For more details, see \cite{gerasimov_penalization_2019,de_lorenzis_nucleation_2022, pham_issues_2011}.

From Griffith's theory, the phase-field approach to brittle fracture inherits the inability to take into account fatigue
effects observed under cyclic loading: 
the crack length $a$ can only grow once the energy release rate $\mathcal{G}$ reaches a critical value $\mathcal{G}_{\text{c}}$, while fatigue-induced crack nucleation and growth occurs at sub-critical loads. In order to incorporate fatigue effects within a variational framework, a fundamental possibility would be to transition to cohesive zone modeling, as advocated in \cite{marigo_modelling_2023}. Thus far, most of the available investigations depart from the variational framework and develop ad hoc extensions of the classical phase-field model to account for fatigue phenomena.
A popular strategy is to gradually decrease the critical energy release rate $\mathcal{G}_{\text{c}}$, also known as fracture toughness, as some representative fatigue history variable accumulates, (see  \cite{alessi_phenomenological_2018,carrara_framework_2020,khalil_generalised_2022,ulloa_phase-field_2021,seles_general_2021,hasan_phase-field_2021} and the overview in \cite{alessi_endowing_2023}).
An alternative strategy is to introduce an additional dissipative energy contribution accounting for fatigue, which leads the damage driving force to increase as some representative fatigue history variable accumulates, as done e.g. in \cite{boldrini_non-isothermal_2016,amendola_thermomechanics_2016, caputo_damage_2015,haveroth_non-isothermal_2020}.
Phase-field fatigue models following other strategies exist, e.g. the one presented in \cite{aygun_coupling_2021} which relies on a cyclic plasticity model, or the one presented in \cite{lo_phase-field_2019} which controls fatigue crack growth by a viscous parameter.
A comprehensive review of phase-field fatigue models is given in \cite{kalina_overview_2023}.

Another classification criterion for phase-field fatigue models is the temporal domain in which the model is formulated:
for the formulation in the physical (pseudo-) time domain $t$, the loading history is to be resolved with several steps per cycle following the path of repetitive loading and unloading.
Alternatively, the model can also be formulated directly in the cycle domain where a surrogate loading sequence replaces the individual steps within a cycle, allowing for an efficient computation of several representative cycles at once, such as e.g. in \cite{seiler_efficient_2020,grossman-ponemon_phase-field_2022,kristensen_accelerated_2023}.
This does not alleviate the inherent history-dependency introduced with the fatigue effects, accompanied by the non-linearity in the cycle domain, implying that for both modeling strategies the computational cost is high.

\begin{figure}
    \centering
    \subfloat[]{
        \includegraphics{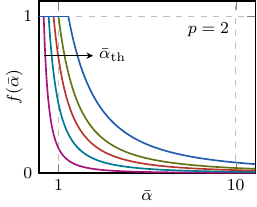}
        \label{fig:fatigue_degradation_ath}
    }
    \subfloat[]{
        \includegraphics{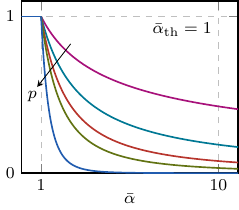}
        \label{fig:fatigue_degradation_p}
    }
    \caption{Fatigue degradation function $f(\bar{\alpha})$ with the influence of its parameters $\bar{\alpha}_{\text{th}}$ (a) and $p$ (b).}
    \label{fig:fatigue_degradation}
\end{figure}

To develop and benchmark our acceleration scheme for phase-field fatigue computations, we adopt the model in \cite{carrara_framework_2020} (extending the one-dimensional model in \cite{alessi_phenomenological_2018}), which served as the basis for many modifications and extensions, e.g. \cite{khalil_generalised_2022,ulloa_phase-field_2021,tan_phase_2022,seles_general_2021,seles_microcrack_2021,hasan_phase-field_2021}.
The core idea is to introduce a \textit{fatigue history variable} $\bar{\alpha}$ to capture the locally endured fatigue effect.
In the mean-load independent version of the model, the fatigue history variable is defined as \cite{carrara_framework_2020}
\begin{equation}
    \bar{\alpha} (\bs{x}; t) = \int_0^t H \left(\alpha \dot{\alpha}\right) |\dot{\alpha}| \dt \text{ ,}
    \label{eq:fat_hist}
\end{equation}
that is the temporal accumulation of a local variable $\alpha (\bs{x})$ representing the `local fatigue effect'.
The Heaviside function $H$ ensures that the fatigue effects are only accumulated during loading phases.
The variable $\alpha$ is defined as the degraded active part of the elastic energy density $g(d) \psi^{\text{el}}_+$.
With this definition, the fatigue effect evolves faster around the crack tip. 
The fatigue history variable enters the so-called \textit{fatigue degradation function} $f(\bar{\alpha}) \in [0,1]$ to modulate the fracture toughness of the material $\mathcal{G}_{\text{c}}$.
It is defined as
\begin{equation}
    f (\bar{\alpha}) =
    \begin{cases}
        1                                                                                                          & \bar{\alpha} < \bar{\alpha}_{\text{th}}    \\
        \left(1 - \frac{\bar{\alpha} - \bar{\alpha}_{\text{th}}}{\bar{\alpha} + \bar{\alpha}_{\text{th}}}\right)^p & \bar{\alpha} \geq \bar{\alpha}_{\text{th}} \\
    \end{cases} \text{ .}
    \label{eq:fatigue_deg_fun_generalized}
\end{equation}
This is slightly different from the one proposed in \cite{carrara_framework_2020}, and features two parameters $p$ and $\bar{\alpha}_{\text{th}}$ that can be calibrated using a limited set of experimental data (e.g., Wöhler and crack growth rate curves) and that allow more flexibility in reproducing different behaviors.
The latter acts as a threshold for $\bar{\alpha}$, governing the point at which the fatigue effect is triggered for the first time, while $p$ influences the rate of degradation, as illustrated in Fig. \ref{fig:fatigue_degradation}.

With these definitions, the rate of the total energy functional for the brittle fatigue fracture model reads
\begin{equation}
    \begin{aligned}
        &\dot{\mathcal{E}} (\bs{u}, d; \bar{\alpha}) =\\
        &\hspace{0.5cm}\hphantom{+} \int_\Omega g(d) \frac{\partial \psi^{\text{el}}_+}{\partial \bs{\varepsilon}}\!:\!\dot{\bs{\varepsilon}} + \frac{\partial \psi^{\text{el}}_-}{\partial \bs{\varepsilon}}\!:\!\dot{\bs{\varepsilon}} +  g^\prime(d) \psi^{\text{el}}_+ \dot{d} \, \mathrm{d}V\\
        &\hspace{0.5cm}+ \int_\Omega f (\bar{\alpha}) \frac{\mathcal{G}_{\text{c}}}{ c_w} \left( \frac{w^\prime (d)}{\ell} \dot{d} \! + 2 \ell \grad d \cdot \grad \dot{d} \right) \mathrm{d}V\\
        &\hspace{0.5cm}- \int_{\partial \Omega_{\text{N}}} \!\!\! \bar{\bs{t}} \cdot \dot{\bs{u}} \mathrm{d}A \, \text{ .}
    \end{aligned}
    \label{eq:energy_functional}
\end{equation}

Departing from a fully variational formulation, $\bar{\alpha}$ is treated as a parameter of the energy functional to derive the governing equations of the coupled field problem.
The balance of linear momentum and the Neumann and Dirichlet boundary conditions read
\begin{equation}
    \begin{aligned}
        \grad \cdot \bs{\sigma} = \bs{0} \qquad &\forall \bs{x} \in \Omega\\
          \bs{\sigma} \cdot \bs{n} = \bar{\bs{t}} \qquad &\forall \bs{x} \in \partial \Omega_{\text{N}}\\
        \bs{u} = \bar{\bs{u}} \qquad &\forall \bs{x} \in \partial \Omega_{\text{D}}
    \end{aligned}
    \label{eq:gov_u}
\end{equation}
with the Cauchy stress definition $\bs{\sigma} := g(d) \partial  \psi^{\text{el}}_+ / \partial \bs{\varepsilon} + \partial \psi^{\text{el}}_- / \partial \bs{\varepsilon}$. Here $\bs{n}$ is the outward  normal unit vector  to the boundary and $\partial\Omega_{\text{D}}$ is the Dirichlet boundary on which $\bar{\bs{u}}$ is the prescribed displacement.
The evolution equations and boundary conditions for the damage variable, due to the postulated irreversibility of damage, are expressed as Karush-Kuhn-Tucker conditions as follows
\begin{subequations}
    \begin{equation}
        \begin{cases}
            \dot{d} \geq 0\\
            \begin{aligned}
                2 (1-d) \psi^{\text{el}}_{+} - \frac{f (\bar{\alpha})\mathcal{G}_{\text{c}}}{c_w} \! \left(\frac{w'(d)}{\ell} - 2 \ell \grad \! \cdot \! (\grad d) \right)\\+ \frac{2 \mathcal{G}_{\text{c}} \ell}{ c_w} \grad f (\bar{\alpha}) \cdot \grad d \leq 0
            \end{aligned}\\
            \begin{aligned}
                \Biggl[ 2 (1-d) \psi^{\text{el}}_{+} - \frac{f (\bar{\alpha})\mathcal{G}_{\text{c}}}{c_w} \! \left(\frac{w'(d)}{\ell} - 2 \ell \grad \! \cdot \! (\grad d) \right)\\+ \frac{2 \mathcal{G}_{\text{c}} \ell}{ c_w} \grad f (\bar{\alpha}) \! \cdot \! (\grad d) \Biggr] \dot{d} = 0
            \end{aligned}\\
        \end{cases}
        \forall \bs{x} \in \Omega
        \label{eq:pf_KKT}
    \end{equation}
    \begin{equation}
        \begin{cases}
            \dot{d} \geq 0\\
            \grad d \cdot \bs{n} \geq 0\\
            (\grad d \cdot \bs{n}) \dot{d} = 0\\
        \end{cases}
        \quad \forall \bs{x} \in \partial \Omega \text{ .}
    \end{equation}
    \label{eq:gov_d}
\end{subequations}
Evolution of the phase-field is only possible once the crack driving force has reached a critical level, which however now depends on the loading history by means of $f(\bar{\alpha})$.

\section{Adaptive acceleration scheme based on a cycle-jump criterion}\label{sec:adaptive_acceleration_based_on_cycle_jump_constraint}
This section addresses the issue of the high  cost associated with phase-field fatigue computations by proposing a novel acceleration scheme.

\subsection{Outline of the scheme}
\begin{figure*}
    \centering
    \includegraphics{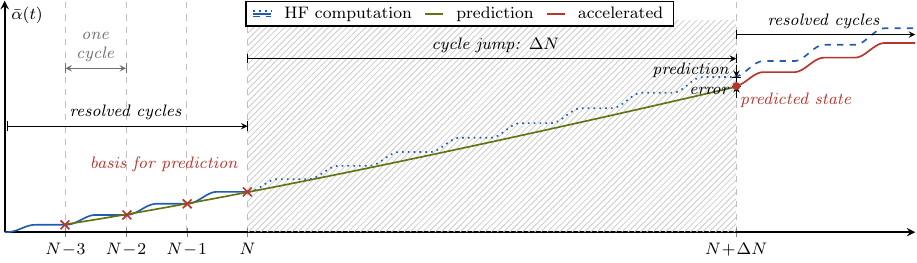}
    \caption{Illustrative representation of the cycle jump technique in terms of the evolution of the fatigue history variable $\bar{\alpha}$ and comparison with the HF computations. Several cycles are computed and used as  basis for the extrapolation of the system evolution over $\Delta N$ cycles. Then, the prediction is enforced and the procedure can start over.}
    \label{fig:cyclejump_notion}
\end{figure*}
We adopt the cycle jump technique due to its straightforward implementation, minimal additional computational cost and general applicability to any type of model formulation.
As schematically illustrated in Fig. \ref{fig:cyclejump_notion}, the strategy is to  resolve some cycles (i.e. compute an appropriate number of load steps during these cycles, as explained later, these are also denoted as high-fidelity or HF computations), extrapolate the further evolution of the system, then jump $\Delta N$ cycles, then again resolve some cycles, and so forth.
Thus, the two core ingredients of the scheme are
\begin{enumerate}
    \item[(i)] the choice of a \textit{local} state variable to be extrapolated to predict the state after the jumped cycles, as well as a strategy to extrapolate it, and
    \item[(ii)] the choice of a \textit{global} state variable and of a criterion based on which to decide whether and, if so, how many cycles can be jumped.
\end{enumerate}
\subsection{Extrapolation of the local system state}
As for the first ingredient, we select the fatigue history variable $\bar{\alpha}$. The starting point for the extrapolation of $\bar{\alpha}$ in the cycle domain is a Taylor series expansion at the last resolved cycle preceding a cycle jump up to the quadratic term to account for non-linear system evolution. It reads
\begin{equation}
    \begin{aligned}
        \bar{\alpha} (\bs{x}, N+\Delta N)
        &= \bar{\alpha} (\bs{x}, N) + \dn{\bar{\alpha}} (\bs{x}, N) \Delta N\\
        &\hphantom{=} + \ddn{\bar{\alpha}} (\bs{x}, N) \frac{\Delta N^2}{2} + \mathcal{O} \left( \Delta N^3 \right) \text{ ,}
    \end{aligned}
    \label{eq:extrapolation_taylor}
\end{equation}
where $\dn{\bar{\alpha}} = \partial\bar{\alpha} /\partial N$ and $\ddn{\bar{\alpha}} := \partial^2\bar{\alpha} /\partial N^2$.
These cycle-wise derivatives are approximated using backwards finite differences (FDs), and we consider a stencil of $N_{\text{s}}=4$ cycles to be computed between cycle jumps, resulting in
\begin{equation}
    \begin{aligned}
        &\bar{\alpha} (\bs{x}, N\!+\!\Delta N) \approx \bar{\alpha}^\star (\bs{x}, N\!+\!\Delta N) = \bar{\alpha} (\bs{x}, N)\\
        &\quad+ \frac{1}{6} \bigl[-2\bar{\alpha} (\bs{x}, N\!-\!3) + 9\bar{\alpha} (\bs{x}, N\!-\!2)\\
        &\qquad\qquad - 18\bar{\alpha} (\bs{x}, N\!-\!1) + 11\bar{\alpha} (\bs{x}, N) \bigr] \Delta N\\
        &\quad+ \frac{1}{2} \bigl[ -\bar{\alpha} (\bs{x}, N\!-\!3) + 4\bar{\alpha} (\bs{x}, N\!-\!2)\\
        &\qquad\qquad -  5\bar{\alpha} (\bs{x}, N\!-\!1) +  2\bar{\alpha} (\bs{x}, N) \bigr] \Delta N^2 \text{ .}\\
    \end{aligned}
    \label{eq:FD_extrapolation}
\end{equation}
Since the state variable to be predicted is a local quantity, the extrapolation must be performed for each  point $\bs{x}$. However, the FD-based scheme has a limited computational cost.

\begin{figure*}
    \centering
    \includegraphics{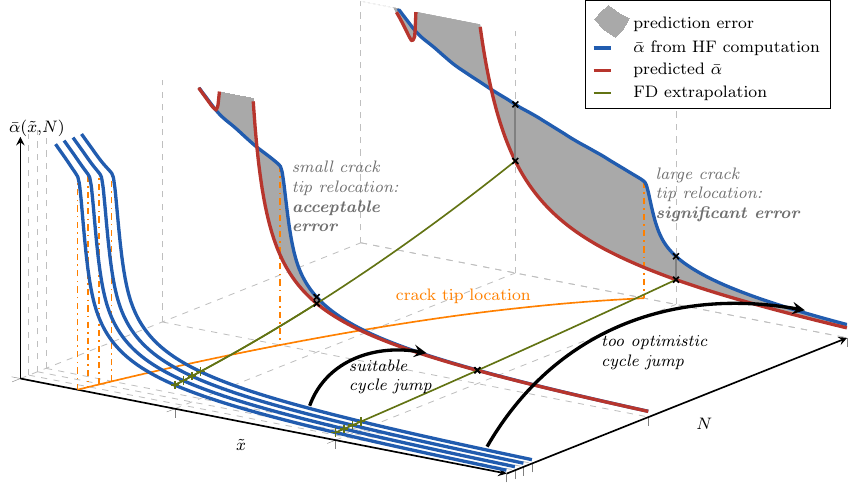}
    \caption{Accuracy of the FD-based extrapolation (\ref{eq:FD_extrapolation}) in comparison to HF computations for two different cycle jump extensions. The blue lines depict the profile of the fatigue history variable $\bar{\alpha}$ ahead of the crack tip along the crack path coordinate $\tilde{x}$.
    A suitable cycle jump with a small crack tip relocation yields a smaller prediction error (gray surfaces) compared to an overly optimistic jump with a large crack tip relocation during the skipped cycles.}
    \label{fig:cyclejump_stageIII}
\end{figure*}
\subsection{Cycle-jump criterion}
In the following, we expand on the second ingredient, i.e. the cycle-jump criterion based on the value of a global variable.
\subsubsection{Need for a cycle-jump criterion}
We schematically motivate the adaptive determination of $\Delta N$ in Fig. \ref{fig:cyclejump_stageIII}. For simplicity, we assume a single fatigue crack propagating stably along the $\tilde{x}$ direction and illustrate the qualitative trend 
of the fatigue history variable $\bar{\alpha}$ along $\tilde{x}$ as a function of the cycle count. $\bar{\alpha}$ is maximum at the crack tip and gradually decreases away from it, and the location of the crack tip changes from cycle to cycle.
If the crack tip position after a cycle jump is sufficiently close to the one prior to it, the local extrapolation of $\bar{\alpha}$ yields acceptable prediction errors, such as for the smaller cycle jump in Fig. \ref{fig:cyclejump_stageIII}.
In contrast, if the crack tip position after a jump is far from the one before the jump, the extrapolation is not accurate, resulting in a large prediction error, such as for the larger cycle jump in Fig. \ref{fig:cyclejump_stageIII}.
Hence, the cycle-jump size $\Delta N$ during crack propagation must be chosen such that the relocation of the crack tip lies within certain bounds.

The generalization of this concept to all stages of the fatigue life is the simple idea behind the proposed ACJ: determine $\Delta N$ such that, during the jumped cycles, the state of the system is only advancing to a predefined extent.
In the following, we formalize the above idea.

\subsubsection{Computation of the number of jumped cycles}
\begin{figure}
    \centering
    \includegraphics{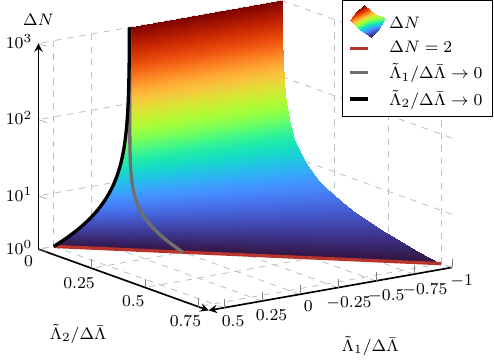}
    \caption{Obtained cycle-jump size $\Delta N$ for a range of the normalized fitting parameters $\tilde{\Lambda}_1/\Delta \bar{\Lambda}$ and $\tilde{\Lambda}_2/\Delta \bar{\Lambda}$. The red line limits the parameter range where the minimum cycle jump size $\Delta N=2$ is obtained, and the black and gray lines represent a purely linear and a purely quadratic fit, respectively.}
    \label{fig:dn_space}
\end{figure}

We define a \textit{global} variable $\Lambda(N)$ which is representative of the current system state from the standpoint of fatigue; the selection of an appropriate variable is discussed in the next subsection.
We then introduce a cycle-jump criterion
\begin{equation}
    \underbrace{\Lambda (\bar{N})}_{\text{post-jump}} \overset{!}{=} \underbrace{\Lambda (N)}_{\text{pre-jump}} + \underbrace{\Delta \bar{\Lambda}\vphantom{(N)}}_{\text{target increment}}
    \label{eq:cycle_jump_criterion_abs}
\end{equation}
imposing a target increment $\Delta \bar{\Lambda}$ of $\Lambda$ during a cycle jump.
Thus, we aim at computing the cycle count $\bar{N} := N + \Delta N$ at which the representative variable takes the value $\Lambda (N) + \Delta \bar{\Lambda}$.
By approximating the left-hand side with a quadratic function 
\begin{equation}
    \Lambda (\bar{N}) \approx \Lambda^\star (\bar{N}) = \tilde{\Lambda}_2 \bar{N}^2 + \tilde{\Lambda}_1 \bar{N} + \tilde{\Lambda}_0,
    \label{eq:Lambda_approx_quad}
\end{equation}
where $\tilde{\Lambda}_0$, $\tilde{\Lambda}_1$ and $\tilde{\Lambda}_2$ are fitting parameters, the cycle-jump criterion can be rewritten as
\begin{equation}
    \tilde{\Lambda}_2 \bar{N}^2 + \tilde{\Lambda}_1 \bar{N} + \tilde{\Lambda}_0 - \Lambda (N) - \Delta \bar{\Lambda} = 0 \text{ .}
    \label{eq:cycle_jump_criterion_approx}
\end{equation}
Based on (\ref{eq:cycle_jump_criterion_approx}), an explicit expression can be found for $\bar{N}$ and hence for $\Delta N$,
\begin{equation}
    \Delta N\!=\!\text{round}\!\left(\!\frac{- \tilde{\Lambda}_1\!+\!\sqrt{ \tilde{\Lambda}_1^2\!-\!4 \tilde{\Lambda}_2\!\left(\!\tilde{\Lambda}_0\!-\!\Lambda (N)\!-\!\Delta \bar{\Lambda} \right) }}{2 \tilde{\Lambda}_2}\!\right) - N \text{ ,}
    \label{eq:Dn_adaptive}
\end{equation}
where we have assumed that $\Lambda^\star$ is monotonically increasing for $\bar{N} > 0$.
This is ensured by choosing $\Lambda$ such that it only grows during the fatigue life, as we explain later.
Further, since the cycle count is an integer, the determined value must be rounded to the nearest natural number.
The obtained cycle-jump size as a function of the normalized fitting parameters $\tilde{\Lambda}_1 / \Delta \bar{\Lambda}$ and $\tilde{\Lambda}_2 / \Delta \bar{\Lambda}$ is visualized in Fig. \ref{fig:dn_space}.
Note that for illustrative purposes and without loss of generality, in Fig. \ref{fig:dn_space}, $\tilde{\Lambda}_0 = \Lambda (N)$ is assumed (meaning that the quadratic approximation matches the intercept of $\Lambda$) and the term $-N$ in (\ref{eq:Dn_adaptive}) is neglected.
We can observe that the cycle-jump size rapidly grows as the fitting parameters tend towards zero, since the slower the evolution of $\Lambda$, the more cycles are necessary to obtain the target increment $\Delta \bar{\Lambda}$.
This manifests the adaptivity of the acceleration scheme.

\begin{figure}
    \centering
    \includegraphics{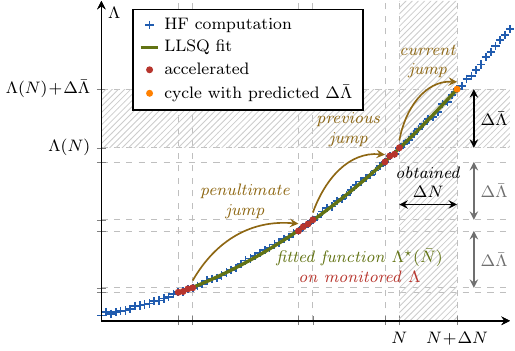}
    \caption{LLSQ fit of the evolution of the representative global state variable $\Lambda$ and cycle jump size $\Delta N$ determined based on the target increment $\Delta\bar{\Lambda}$.}
    \label{fig:cyclejump_LLSQ}
\end{figure}

To obtain the fitting parameters of the quadratic function $\tilde{\Lambda}_{0,1,2}$, we use a linear least squares (LLSQ) fit.
To ensure sufficient robustness in the computation of $\Delta N$, the quadratic function is fitted onto data points of $\Lambda$ during the last $3 N_\text{s}$  computed cycles, i.e. ranging up to the penultimate cycle jump.
This is especially crucial for the stability of the acceleration scheme in the HCF regime, where $\Lambda$ may evolve very slowly.
Since only a short segment of the evolution of $\Lambda$ is represented by the quadratic interpolation (\ref{eq:Lambda_approx_quad}), the fitting parameters $\tilde{\Lambda}_{0,1,2}$ must be re-determined prior to any individual cycle jump.
The LLSQ fit on the data points between three consecutive cycle jumps and the general concept of the outlined approach are illustrated in Fig. \ref{fig:cyclejump_LLSQ}, where the various cycle jumps lead to the same $\Delta \bar{\Lambda}$, although achieved with different cycle-jump sizes.

\subsubsection{Automatic transition to high-fidelity computation}\label{sec:automatic_return_to_hf_computation}

It is worth noting that the fatigue computations are only accelerated if \eqref{eq:Dn_adaptive} delivers $\Delta N \geq 2$, which by virtue of (\ref{eq:cycle_jump_criterion_approx}) can be translated into feasible ranges of the fitting parameters $\tilde{\Lambda}_{1}$ and $\tilde{\Lambda}_2$, namely
\begin{equation}
    4 \tilde{\Lambda}_2 + 2 \tilde{\Lambda}_1 \leq \Delta \bar{\Lambda} \text{ ,}
\end{equation}
again assuming $\tilde{\Lambda}_0 = \Lambda (N)$ and neglecting the term $-N$ in (\ref{eq:Dn_adaptive}) for illustrative purposes without loss of generality.
This is drawn as the red limit curve in Fig. \ref{fig:dn_space}.
The minimum number of cycles to be jumped corresponds to the rate of system evolution being so large that the acceleration scheme naturally reverts to a HF analysis.
I.e., once the algorithm determines that the evolution of the system is too fast  by obtaining $\Delta N < 2$, no jump is performed.
At very low load levels, this transition to the full resolution of the remaining fatigue life may occur very late within the fatigue life or not at all; for higher load levels, it may happen relatively early, far from failure.
Further, at load levels close to monotonic failure, the algorithm  does not skip cycles since the system evolution is too fast.

\subsubsection{Special cases}\label{sec:special_cases}
Despite the choice of a monotonically increasing $\Lambda$, in a numerical context with the fitted function $\Lambda^\star$, two special cases must be addressed.
The first special case occurs if  (\ref{eq:Dn_adaptive}) does not have a real solution, i.e.
the parabola described by the quadratic ansatz (\ref{eq:Dn_adaptive}) does not obtain the target value of $\Lambda (N) + \Delta \bar{\Lambda}$ for any real $\Delta N$.
In this case, we revert to a linear approximation of the evolution of $\Lambda$, i.e. we neglect the quadratic term and compute a cycle-jump size of
\begin{equation}
    \begin{aligned}
        &\Delta N = \text{round} \left( \frac{\Lambda(N) + \Delta \bar{\Lambda} - \tilde{\Lambda}_{0\text{,lin}}}{\tilde{\Lambda}_{1\text{,lin}}} \right) - N\\
        &\qquad\text{if} \quad \tilde{\Lambda}_1^2 - 4 \tilde{\Lambda}_2\!\left(\!\tilde{\Lambda}_0\!-\!\Lambda (N)\!-\!\Delta \bar{\Lambda} \right) < 0 \text{ .}
    \end{aligned}
\end{equation}
Clearly, the fitting parameters $\tilde{\Lambda}_{p\text{,lin}}$ need to be re-determined.
The second special case occurs if (\ref{eq:Dn_adaptive}) yields a negative cycle-jump size.
This may occur in the VHCF regime with extremely slow system evolution. In this case, we simply use half of $\Delta N$ of the last successfully performed cycle jump.

\subsection{Choice of the global monitoring variable and of its target increment}\label{sec:definition_of_control_quantities}
\begin{figure*}
    \centering
    \includegraphics{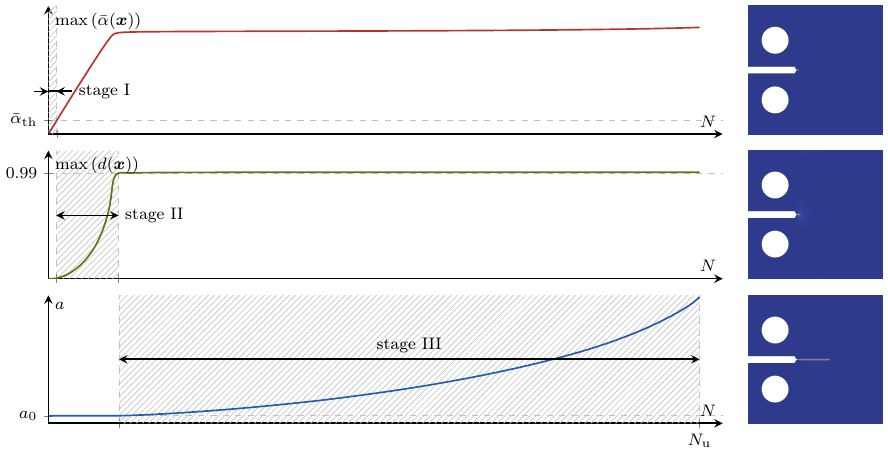}
    \caption{Stages I, II and III of the fatigue lifetime for a phase-field fatigue model in terms of the monitored global state quantities $\max (\bar{\alpha})$, $\max (d)$ and crack length $a$. On the right-hand side, exemplary phase-field states from a representative test are depicted.}
    \label{fig:cyclejump_stages}
\end{figure*}

For the proposed strategy, the choice of $\Lambda$ and $\Delta \bar{\Lambda}$ plays a central role.
Further, $\Lambda$ must be monotonically increasing as the fatigue effects advance.
Due to the diversity of the system behavior in different phases of its fatigue life, no single choice for $\Lambda$ is appropriate for all phases. Thus, we select a different $\Lambda$ in each of three fatigue life stages, as depicted in Fig. \ref{fig:cyclejump_stages} and described in the following.

\subsubsection{Stage I: pre-fatigue effect}
\label{sec:stageI_pre_fatigue_impact}
\begin{figure}
    \centering
    \subfloat[]{
        \includegraphics{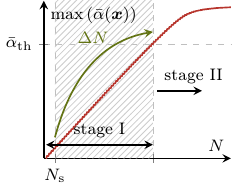}
        \vspace*{-0.25cm}
        \label{fig:cyclejump_stageI_fathist}
    }
    \subfloat[]{
        \includegraphics{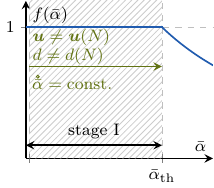}
        \vspace*{-0.25cm}
        \label{fig:cyclejump_stageI_fatdeg}
    }
    \caption{Evolution of $\max (\bar{\alpha})$ in the first stage (a) and behavior of the fatigue degradation function $f(\bar{\alpha})$ (b).}
    \label{fig:cyclejump_stageI}
\end{figure}

As outlined in Section \ref{sec:phase-field_fatigue_modeling}, the adopted phase-field fatigue model features a threshold value $\bar{\alpha}_{\text{th}}$ of the fatigue history variable, prior to which no fatigue effects take place.
Thus, this initial stage clearly satisfies
\begin{equation}
    \underset{\bs{x}}{\max} \left( \bar{\alpha} (\bs{x}) \right) \leq \bar{\alpha}_{\text{th}} \text{ .}
\end{equation}
As depicted in Fig. \ref{fig:cyclejump_stageI_fathist}, the fatigue history variable $\bar{\alpha}$ accumulates linearly from cycle to cycle during stage I, since $\bar{\alpha}<\bar{\alpha}_{\text{th}}$.
Consequently, the system behaves linear-elastically and it is possible to jump directly to the point where the fatigue effect is triggered for the first time, which can be determined analytically due to the linear growth.
This means that for the first stage, the choice for $\Lambda$ is $\max \left( \bar{\alpha} (\bs{x}) \right)$ while the target increment $\Delta \bar{\Lambda}$ is
\begin{equation}
    \Delta \max \left( \bar{\alpha} (\bs{x}) \right) = \bar{\alpha}_{\text{th}} - \max \left( \bar{\alpha} (\bs{x}) \right) \text{ .}
\end{equation}
Essentially, this makes for an initial cycle jump similar to the one presented in \cite{seles_general_2021} where the cycle regime without fatigue effects is skipped directly.
As indicated in Fig. \ref{fig:cyclejump_stageI_fathist}, the first $N_{\text{s}}$ cycles must be resolved to obtain a basis for the LLSQ fit.
The achievable speed up in stage I depends on the chosen fatigue history variable threshold $\bar{\alpha}_{\text{th}}$ as well as the loading; in general however, stage I amounts to a rather small portion of the overall fatigue lifetime.

\subsubsection{Stage II: fatigue crack nucleation}
\begin{figure}
    \centering
    \hspace*{-0.5cm}\includegraphics{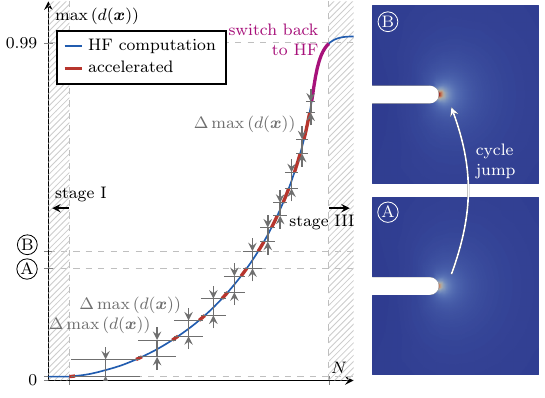}
    \caption{Comparison between the HF and the ACJ evolution of $\max (d)$ in the second stage of the fatigue life (left) and evolution of the phase-field variable at two representative points \textsf{A} and \textsf{B}. The ACJ criterion is defined so as to  trigger a constant increment of $\Delta \max(d)$ during a cycle jump.}
    \label{fig:cyclejump_stageII}
\end{figure}
Once the fatigue threshold is met, fatigue effects start. The damage variable, initially zero, evolves until at some location the first fully developed damage localization profile (the smeared representation of a crack) appears, i.e. at some point it is $d> 1-\mathtt{TOL}$, where we choose $\mathtt{TOL} = 0.01$. Thus, $\Lambda$ is $\max(d(\boldsymbol{x}))$, and the second stage is identified by
\begin{equation}
    \underset{\bs{x}}{\max} \left( d (\bs{x}) \right) \leq 0.99 \text{ .}
\end{equation}
The stage-II cycle jump splits this range by imposing equal increments of $\Delta \max (d)$, as illustrated in Fig. \ref{fig:cyclejump_stageII}.
The largest cycle jumps are obtained in the beginning of the stage with relatively slow system evolution, while towards the end of the stage, $\Delta N$ becomes smaller.
If $\max(d)$ evolves too quickly, as illustrated in Fig. \ref{fig:cyclejump_stageII}, the determined cycle-jump size is $\Delta N < 2$, hence no further cycles are skipped and the remaining cycles are resolved.
We propose a target increment $\Delta \bar{\Lambda}$ of $0.02$ in this second stage.
To provide flexibility, a scalar factor $\lambda_{\text{II}} > 0$ is introduced to influence this baseline, i.e.
\begin{equation}
    \Delta \max \left( d (\bs{x}) \right) = \lambda_{\text{II}} 0.02 \text{ .}
\end{equation}
This means that for $\lambda_{\text{II}} > 1$, the target increment during the cycle jump is larger than the baseline, obtaining larger $\Delta N$ which increase the speedup but decrease the accuracy of the fatigue computations.
Conversely, for $\lambda_{\text{II}} < 1$, the accuracy improves with smaller cycle jumps, while the speedup decreases.
With the default choice of $\lambda_{\text{II}} = 1$, the acceleration scheme in our numerical experiments always yields a good compromise between speedup and accuracy, independently of the specimen, material or load level, as shown in Section \ref{sec:numerical_examples}.

\subsubsection{Stage III: fatigue crack propagation}\label{sec:stage_III_fatigue_crack_propagation}
\begin{figure*}
    \centering
    \includegraphics{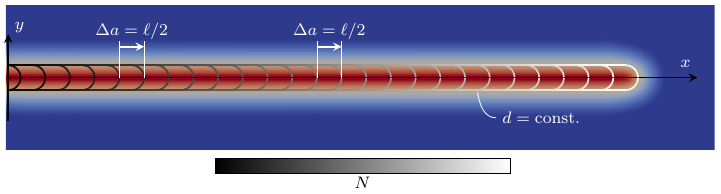}
    \caption{Discretization of the phase-field crack by means of cycle jumps with a constant crack increment $\Delta a$ in stage III. The contour lines of the phase field indicate different crack lengths in between the cycle jumps.}
    \label{fig:cyclejump_stageIII_a}
\end{figure*}

After the formation of the first crack, the crack grows, first stably, and finally unstably, until failure at $N_{\text{u}}$.
In this stage, the crack length  itself, $a$, is the most representative monitoring quantity for $\Lambda$ which satisfies
\begin{equation}
     a \leq a_{\text{u}} \text{,}
\end{equation}
with $a_{\text{u}}$ as the crack length at failure.
To fix ideas, we consider here a single fatigue crack, while the approach is able to account for multiple fatigue cracks propagating at the same time, see Section \ref{sec:numerically_efficient_crack_tip_tracking}. As will be shown with the numerical examples in Section \ref{sec:numerical_examples}, most cycle jumps occur in this stage during stable crack propagation.
In stage III, the target increment $\Delta a$ must be associated to the size of the zone with dominant fatigue effect.
This is why we define the feasible increment $\Delta \bar{\Lambda}$ in stage III as
\begin{equation}
    \Delta a = \lambda_{\text{III}} \frac{\ell}{2} \text{ .}
\end{equation}
Analogously to stage II, a parameter $\lambda_{\text{III}}>0$ is introduced for more flexibility.
Again, $\lambda_{\text{III}}>1$ allows for larger crack propagation during a cycle jump resulting in higher speedup and lower accuracy, and vice versa for $\lambda_{\text{III}}<1$.
The cycle jumps of stage III can be interpreted as a subdivision of the total crack length into multiple jumps with uniform crack increments, as portrayed in Fig. \ref{fig:cyclejump_stageIII_a}.
Due to the ever increasing crack growth rate $\dn{a}$ for a standard fatigue test setup (Fig. \ref{fig:cgr_adaptivity}), also in stage III the cycle-jump size decreases during the progression of the fatigue life.

The choice of the crack length $a$ as representative variable within stage III raises the challenge of how to precisely monitor the crack length in a numerical context.
Conventional crack tip tracking algorithms monitor no crack growth below the discretization length, which would cause the proposed scheme to fail in the HCF regime.
To overcome this and further issues, we introduce the notion of the `smeared crack length' in Section \ref{sec:numerically_efficient_crack_tip_tracking}.

\begin{table*}
    \caption{Stage-wise characteristics of the ACJ scheme.}
    \label{tab:stage_wise_acceleration_overview}
    \centering
    \footnotesize
    \begin{tabularx}{0.95\textwidth}{lllll}
        \toprule
        stage &stage identification &$\Lambda$ &$\Delta \bar{\Lambda}$ &comments\\
        \midrule
        I
              &$\max \left( \bar{\alpha} (\bs{x}) \right) \leq \bar{\alpha}_{\text{th}}$
              &$\max \left( \bar{\alpha} (\bs{x}) \right)$
              &$\bar{\alpha}_{\text{th}} - \max \left( \bar{\alpha} (\bs{x}) \right)$
              &linear evolution after first full cycle for chosen model \cite{carrara_framework_2020}\\
        II
              &$\max \left( d (\bs{x}) \right) \leq 0.99$
              &$\max \left( d (\bs{x}) \right)$
              &$\lambda_{\text{II}} 0.02$
              &\\
        III
              &$\max \left( d (\bs{x}) \right) > 0.99$
              &$a$
              &$\lambda_{\text{III}} \ell/2$
              &smeared crack length for crack tip tracking (Section \ref{sec:smeared_crack_length_approach})\\
        \bottomrule
    \end{tabularx}
\end{table*}
An overview of the resulting stage-wise acceleration scheme is given in Tab. \ref{tab:stage_wise_acceleration_overview} in terms of stage definition, representative variable $\Lambda$ as well as target increment $\Delta \bar{\Lambda}$ thereof during a jump.
For an external loading which causes $\max \left( \bar{\alpha} (\bs{x}) \right) > \bar{\alpha}_{\text{th}}$ already in the first cycle, the computation directly starts in stage II; if the external loading leads to full crack development already in the first cycle, only stage III remains.

\subsubsection{Estimate of the number of cycles to be resolved}\label{sec:summary_and_estimation_of_N_r}
A noteworthy feature of the proposed approach is that, given the domain geometry and material parameters, nearly the same total number of cycles must be computed for the total fatigue life regardless of the specific rate of system evolution, which we illustrate in the following.
Since the cycle-jump criterion ensures the target increment $\Delta \bar{\Lambda}$ of the global monitoring variable, the number of cycles to be resolved $N_{\text{r}}$ can be estimated as follows
\begin{equation}
    \begin{aligned}
        N_{\text{r,est}} = &\underbrace{N_{\text{s}}\vphantom{\bigg|}}_{\text{stage I}} + \underbrace{\frac{0.99}{\lambda_{\text{II}}0.02} N_{\text{s}}\vphantom{\bigg|}}_{\text{stage II}} + \underbrace{\frac{a_{\text{u}}}{\lambda_{\text{III}}\ell/2} N_{\text{s}}\vphantom{\bigg|}}_{\text{stage III}} \text{ .}
    \end{aligned}
    \label{eq:nresolved_estimation}
\end{equation}
While for the first stage a single cycle jump is necessary, for the other stages the number of cycle jumps originates from the total range of the representative variable divided by the imposed increment.
$N_{\text{s}}$ cycles have to be computed in between the individual cycle jumps, and hence $N_{\text{s}}$ cycles are counted per cycle jump within each stage. Thus, the number of cycles to be resolved only depends on the speedup parameters, $N_s$ and $a_{\text{u}}/\ell$.
As an example, with speedup parameters $\lambda_{\text{II}} = \lambda_{\text{III}} = 1$, $N_s=4$ and assuming  $a_{\text{u}} = 100 \ell$, this results in $1002$ cycles which need to be resolved.
Note that this estimate assumes that the system predominantly evolves during the cycle jumps and exhibits a negligible system evolution within resolved cycles, which is the case for HCF computations; for LCF computations, the system shows a substantial evolution also during the cycles between cycle jumps, meaning that less cycles than estimated need to be resolved.
The validity of this estimate is shown later with numerical examples.
Note that for $\lambda_{\text{II,III}} > 1$, the total number of cycles to be resolved decreases, while for $\lambda_{\text{II,III}} < 1$ the number of resolved cycles increases.
Also, the speedup parameters $\lambda_{\text{II}}$ and $\lambda_{\text{III}}$ allow to influence the accuracy of the stage of crack nucleation or crack propagation independently.
An inverse consideration to (\ref{eq:nresolved_estimation}) can be made, meaning that the speedup parameters can be tuned based on the number of computationally affordable cycles.

\subsection{Algorithmic aspects}\label{sec:algorithmic_aspects}
\begin{figure*}
    \centering
    \hspace{-1cm}
    \subfloat[]{
        \includegraphics{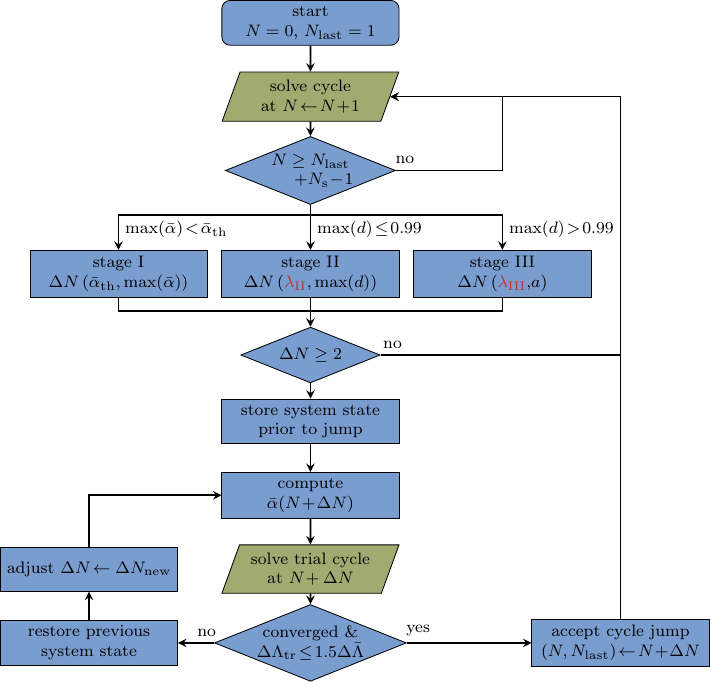}
        \label{fig:cyclejump_procedure}
    }
    \hspace{-1cm}
    \subfloat[]{
        \raisebox{2.375cm}{\includegraphics{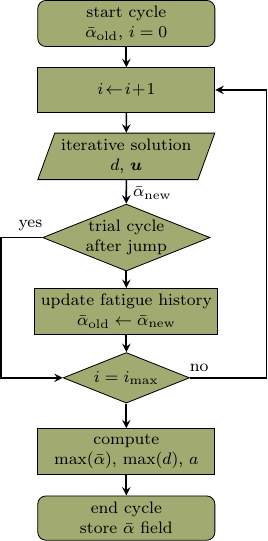}}
        \label{fig:cyclesolver_flowchart}
    }
    \caption{Flowchart illustrating the steps of the acceleration scheme (a) and of the solution of an individual cycle (b).}
    \label{fig:cyclejump_flowchart}
\end{figure*}
To conclude this section, we address algorithmic aspects relevant for the implementation of the scheme.
A flowchart is shown in Fig. \ref{fig:cyclejump_flowchart}.
The procedure starts by resolving a cycle, i.e. solving multiple load steps with an iterative solution procedure, as detailed in Fig. \ref{fig:cyclesolver_flowchart}.
After each resolved cycle, the monitoring variables are computed.
Furthermore, the $\bar{\alpha}$ field is saved for the last $N_{\text{s}}$  computed cycles.

After the end of a cycle, the algorithm checks based on the cycle count whether a cycle jump can be performed:
since the last executed cycle jump to the current cycle number $N_{\text{last}}$, at least $N_{\text{s}}$ cycles must have been computed to have a basis for the FD-based system state prediction.
If this is not the case, the next cycle is resolved again, otherwise, the algorithm determines in which stage of fatigue life the system is  based on the value of the variables $\max\left(\bar{\alpha} (\bs{x})\right)$ and $\max\left(d (\bs{x})\right)$. Then, the extension of the cycle-jump $\Delta N$ is computed using \eqref{eq:Dn_adaptive} and considering the special cases in Section \ref{sec:special_cases}.
For this, the LLSQ fit is performed on the data points of $\Lambda$ from the last $3 N_{\text{s}}$ computed cycles (at the beginning of the computations where less data points are available, all available data points are used, i.e. $1N_{\text{s}}$ or $2N_{\text{s}}$ for the first and second cycle jump, respectively).
If a cycle-jump size $\Delta N < 2$ is obtained, no cycle jump is performed, but the next cycle is again resolved.
In the case of $\Delta N \geq 2$, the system state at $N + \Delta N$ is predicted with the FD-based extrapolation (\ref{eq:FD_extrapolation}).
To enforce the irreversibility of the fatigue process $\dot{\bar{\alpha}}\geq 0$, any negative increments are neglected in the FD-based extrapolation of $\bar{\alpha}$ (such values may occur when the system evolution rate is close to numerical tolerances in the VHCF regime).

After the cycle jump, we compute the cycle number $N + \Delta N$, i.e. the cycle for which the monitoring variable is expected to be $\Lambda(N)+\Delta \bar{\Lambda}$, which we denote as \textit{trial cycle} since it is the cycle at which we decide whether to accept or reject the cycle jump.
This cycle follows exactly the same procedure as a regular resolved cycle illustrated in Fig. \ref{fig:cyclesolver_flowchart}, with the difference that $\bar{\alpha}$ is not updated during the solution procedure, since its value at the end of the trial cycle is already known.
The trial cycle serves several purposes:
\begin{itemize}
    \item By computing the system state at $N + \Delta N$, the equilibrium of the system is re-established with the extrapolated state variable.
    \item Instead of computing only the  last load step of the cycle $N + \Delta N$, the whole trial cycle is resolved before accepting the cycle jump.
    If any of the tested load steps of the trial cycle do not converge, implying that the system equilibrium could not be established with the extrapolated state variables (e.g. due to a too optimistic cycle jump), the state prior to the jump is restored.
    Then, another cycle jump is tried with half of the last accepted cycle-jump size.
    Only if all load steps of the trial cycle converge, the cycle jump is accepted.
    \item Finally, the trial cycle allows to determine the satisfaction of the cycle-jump criterion (\ref{eq:cycle_jump_criterion_abs}) during the skipped cycles, i.e. an \textit{a posteriori} error evaluation.
    For this, the growth $\Delta \Lambda_{\text{tr}}$ determined after the trial cycle  is  compared to the target increment $\Delta \bar{\Lambda}$.
    If the cycle-jump criterion is violated by more than $50 \%$, the old state is restored and the jump size for the next trial is adjusted to
\end{itemize}
\begin{equation}
    \Delta N_{\text{new}} = \text{round} \left( \frac{\Delta \bar{\Lambda}}{\Delta \Lambda_{\text{tr}}} \Delta N \right) \quad \text{if} \quad \Delta \Lambda_{\text{tr}} > 1.5 \Delta \bar{\Lambda} \text{ .}
    \label{eq:correction_cycle_jump}
\end{equation}
The computational effort for the trial cycle is not wasted, since the system state at $N + \Delta N$ can be used as basis for the following cycle jump, hence only $N_{\text{s}}-1$ cycles remain to be computed prior to the next cycle jump.
Note that the adjustment of the cycle-jump size after non-convergence, or non-satisfaction of the cycle-jump criterion is only rarely necessary in the VHCF regime or close to transitions between stages, as demonstrated later with numerical examples.

\subsection{Applicability of the approach}
Although the approach was presented with the model presented in \cite{carrara_framework_2020}, it is applicable to any other phase-field fatigue model.
This is due to the general definition of the three stages of the fatigue lifetime as well as the universal definition of the cycle-jump criterion with increments of $\Lambda$.
For other phase-field fatigue models, solely the specific choice for the variable to be extrapolated as well as the target increments $\Delta \bar{\Lambda}$ may be different.
E.g. for models featuring cyclic plasticity \cite{aygun_coupling_2021,seles_general_2021,ulloa_phase-field_2021}, one choice could be to extrapolate the accumulated plastic strain during the skipped cycles.
Furthermore, for phase-field fatigue models directly formulated in the cycle domain \cite{seiler_efficient_2020,schreiber_phase_2020}, the proposed concept can be used as an adaptive cycle stepping algorithm.

Note also that, the extension to larger problems (e.g. in 3D) should not affect the efficiency of the method since the extra computational effort of the ACJ scales linearly with the problem size. This is related to the fact that only norms of arrays must be tracked to determine the cycle jump size and the explicit FD-based system state prediction can be fully vectorized.

\section{Smeared crack tip tracking}\label{sec:numerically_efficient_crack_tip_tracking}
In stage III, the proposed acceleration scheme relies on an accurate measurement of the crack growth rate  to adaptively determine the cycle-jump size.
In the following, we discuss why conventional crack tip tracking algorithms cannot be used, and propose the smeared crack length concept.

\subsection{Issues of existing crack tip tracking algorithms}
\label{sec:infeasibility_of_conventional_crack_tip_tracking_algorithms}
In a finite element (FE) computation, the simplest form of crack tip tracking is to identify the nodes with a phase-field value above a numerical threshold $d_{\text{th}}$ as broken, and track the maximum (or minimum) of all broken nodal coordinates with respect to a given origin, as outlined e.g. in \cite{hansen-dorr_phase-field_2020}.
This means however that the direction of crack propagation must be known \textit{a priori}, while the minimal detectable crack growth is the discretization length $h$.
This dependency on the spatial discretization is especially problematic for fatigue computations, where the crack growth rate may be significantly smaller than the discretization length, meaning that no crack propagation is observed for possibly thousands of cycles.
More involved crack tip tracking strategies to automatically determine the direction of crack propagation or to incorporate branching could be an option, see e.g.  \cite{zeng_tracking_2022}, however they add computational cost. 

\subsection{Smeared crack length approach}\label{sec:smeared_crack_length_approach}
\begin{figure}
    \centering
    \includegraphics{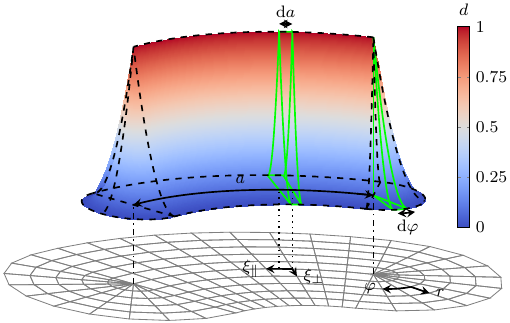}
    \caption{Crack tip and extension contributions of the optimal phase-field integral and local coordinate systems for their integration (exemplarily for the AT1 model).}
    \label{fig:optimal_phasefield_volume}
\end{figure}

The proposed cycle-jump scheme needs a crack tip tracking algorithm capable of detecting very small crack growth, without knowing the crack path \textit{a priori}, while still being computationally efficient.
To achieve this, we compute the crack length $a$ by equating the integral of the numerically obtained phase-field solution $D_{\text{num}} (d)$ with the theoretical value obtained by integrating the optimal phase-field profile $D_{\text{opt}} (a, \ell)$, i.e.
\begin{equation}
    D_{\text{num}} (d) = D_{\text{opt}} (a, \ell) \text{ .}
    \label{eq:D_numopt}
\end{equation}
To obtain $D_{\text{opt}} (a, \ell)$, for a 2D setting, we assume that the optimal phase-field profile is extruded along the crack axis, and that its half-profile is revolved around the crack tip, as illustrated in Fig. \ref{fig:optimal_phasefield_volume} exemplarily for the AT1 model.
Hence, we split $D_{\text{opt}}$ into contributions of the crack length $D_{\text{ext}}$ and of the crack tip $D_{\text{tip}}$,
\begin{equation}
    D_{\text{opt}} (a, \ell) = D_{\text{ext}} (a, \ell) + k D_{\text{tip}} (\ell) \text{ .}
\end{equation}
The factor $k\in\mathbb{N}$ allows to account for multiple crack tip contributions and must be set based on the expected number of crack tips.
With the coordinate definition in Fig. \ref{fig:optimal_phasefield_volume}, the two contributions read \cite{gerasimov_penalization_2019}
\begin{subequations}
    \begin{equation}
        \begin{aligned}
            &D_{\text{ext}} (a, \ell) = \int_{\xi_\parallel=0}^a \int_{\xi_\perp}\!\!\!\!d(\xi_\perp) \mathrm{d}\xi_\perp \mathrm{d}\xi_\parallel = a
            \begin{cases}
                \frac{4}{3} \ell &\text{AT1}\\
                2 \ell           &\text{AT2}\\
            \end{cases}\\
            &\qquad\qquad\qquad\text{and}\\
        \end{aligned}
    \end{equation}
    \begin{equation}
        D_{\text{tip}} (\ell) = \int_{\varphi=0}^{\pi} \int_r\!d(r, \varphi) r \mathrm{d}r \mathrm{d} \varphi = \pi
        \begin{cases}
            \frac{1}{3} \ell^2 &\text{AT1}\\
            \ell^2             &\text{AT2}\\
        \end{cases} \text{ ,}
    \end{equation}
\end{subequations}
Rearranging (\ref{eq:D_numopt}) for $a$ yields
\begin{equation}
    a = \frac{D_{\text{num}} (d) - k D_{\text{tip}}(\ell)}{D_{\text{ext}} (\ell)} =
    \begin{cases}
        \frac{D_{\text{num}}(d) - k \pi \frac{1}{3} \ell^2}{4/3 \ell} & \text{AT1} \\
        \frac{D_{\text{num}}(d) - k \pi \ell^2}{2 \ell}               & \text{AT2} \\
    \end{cases} \text{ .}
    \label{eq:smeared_crack_length}
\end{equation}
and the smeared crack growth rate reads
\begin{equation}
    \dn{a}= 
    \begin{cases}
        \frac{\dn{D}{}_{\text{num}}(d)}{4/3 \ell} & \text{AT1} \\
        \frac{\dn{D}{}_{\text{num}}(d)}{2 \ell}   & \text{AT2} \\
    \end{cases} \text{ .}
\end{equation}
Crack branching or merging, or more generally multiple cracks are automatically dealt with. An extension to the 3D case to compute the crack surface is straightforward.
In a numerical setting, the assumption of an optimal phase-field profile may not hold true due to the irreversibility condition and the discretization error.
Their influence on the smeared crack length is investigated and corrected in Appendix \ref{sec:corrections_of_the_smeared_crack_length}.

In a standard FE implementation, the computation of $D_{\text{num}}$ involves numerical integration.
An efficient integration is possible when assuming that the phase-field support is discretized uniformly.
In this case, one can replace the numerical integration with the $L1$-norm of the phase-field solution vector $\norm{\mathbf{d}}{1}$,
\begin{equation}
    D_{\text{num}} (d) = J w \norm{\mathbf{d}}{1} \qquad \text{if} \qquad h = \text{const.}
    \label{eq:D_num_L1_approx}
\end{equation}
with a constant Jacobian $J$ for the elements in the support, and sum of the numerical integration weights $w$.
Both ways to compute the smeared crack length are compared and validated with conventional crack tip tracking algorithms in Appendix \ref{sec:validation_of_the_smeared_crack_length}.

\section{Numerical examples}\label{sec:numerical_examples}
In this section, the performance of the proposed adaptive acceleration scheme is demonstrated through numerical examples and compared with that of some other available acceleration schemes.

\subsection{Numerical setup}\label{sec:numerical_aspects}
\begin{figure}
    \centering
    \includegraphics{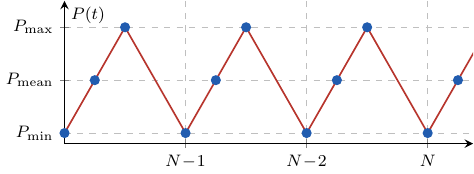}
    \caption{Discretization of the cyclic load applied (red curve) with multiple load steps (blue circles).}
    \label{fig:loading_curve_discretization}
\end{figure}
For the numerical examples in this work, a standard FE procedure with a staggered solution scheme is adopted to solve the coupled problem.
This means that the weak form of eqs.  (\ref{eq:gov_u}) and  (\ref{eq:gov_d}) is solved in an alternate fashion using an iterative Newton-Raphson scheme with a tolerance of $\mathtt{TOL}_{\text{NR}} = 10^{-6}$ for the norm of the residual, until both residual norms fall below $\mathtt{TOL}_{\text{stag}} = 10^{-4}$.
All computations in the following are performed in the 2D setting and results are reported per unit thickness.
The irreversibility of the phase field is enforced using the history-field approach \cite{miehe_phase_2010}, while different local dissipation functions and energy decompositions are adopted to assess the general applicability of the acceleration scheme.
For the AT1 dissipation function, the recovery penalty proposed in \cite{gerasimov_penalization_2019} is used to ensure $d\geq 0$.
Bilinear quadrilateral elements are used, while the meshes are locally refined in the areas of the expected crack.
To discretize the loading cycles, the loading phases are resolved using multiple steps, while the intermediate load steps of the unloading phase are not resolved to save computational time.
This can be done without loss of accuracy since the fatigue history variable only accumulates in the loading phases and the system behaves linearly during the unloading phases.
Nonetheless, the final point of the unloading phase must be resolved such that the following loading phase is captured correctly, and since it is the reference point for the cycle-jump procedure.
In this work, triangular load functions are considered with three load steps per cycle, that is two during loading and one during unloading as highlighted by the blue points in Fig. \ref{fig:loading_curve_discretization}.

\begin{figure}
    \centering
    \subfloat[]{
        \includegraphics{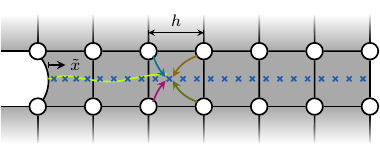}
        \label{fig:cracktip_interpolation_element_interpolation}
        \vspace{-0.25cm}
    }\\[-1em]
    \subfloat[]{
        \includegraphics{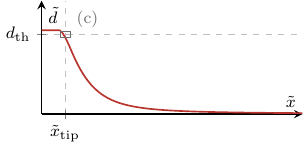}
        \label{fig:cracktip_interpolation_phase-field_interpolation}
    }
    \subfloat[]{
        \hspace{-0.25cm}\includegraphics{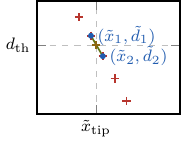}
        \label{fig:cracktip_interpolation_cracktip_interpolation}
    }
    \caption{Steps to obtain the interpolated crack tip position, featuring the element-wise projection of the phase-field solution to the inter-nodal regime (a), the selection of relevant points of the projected phase field along a crack path coordinate $\tilde{d}(\tilde{x})$ (b), and the linear interpolation of the crack tip position $\tilde{x}_{\text{tip}}$ (c).}
    \label{fig:cracktip_interpolation}
\end{figure}

As clarified in Appendix~\ref{sec:validation_of_the_smeared_crack_length}, the adoption of the smeared crack length approach introduces an approximation in the determination of the crack length. To avoid including such effect in the following comparisons, we adopt here a conventional crack tip tracking algorithm.
To still detect crack length increments smaller than the discretization length (Section \ref{sec:infeasibility_of_conventional_crack_tip_tracking_algorithms}), we use an interpolated crack tip approach for specimens where the crack path is known \textit{a priori} to obtain the crack growth rate curves.
As illustrated in Fig. \ref{fig:cracktip_interpolation}, we project the nodal phase-field solution onto a crack path coordinate $\tilde{x}$ with the shape functions of the elements crossed by the crack (Fig. \ref{fig:cracktip_interpolation_element_interpolation}).
Based on the projected phase field along the crack path $\tilde{d}(\tilde{x})$, Fig. \ref{fig:cracktip_interpolation_phase-field_interpolation}, we can determine the position along the crack path where $\tilde{d} (\tilde{x}) = d_{\text{th}}$.
This is done by linearly interpolating the projected phase-field values, as depicted in Fig. \ref{fig:cracktip_interpolation_cracktip_interpolation}.
The crack growth rate is assessed at interpolated crack length increments of $\Delta a = \ell / 2$.

\subsection{Compact tension test}
\label{sec:behavior_of_the_proposed_approach}
\begin{figure}
    \centering
    \subfloat[]{
        \hspace*{-0.5cm}\includegraphics{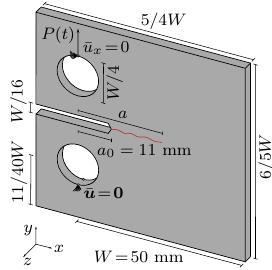}\hspace*{-0.3cm}
        \label{fig:specimen_ct}
    }
    \subfloat[]{
        \begin{tikzpicture}
            \node[anchor=center] at (0,0) {\includegraphics[width=3.509cm]{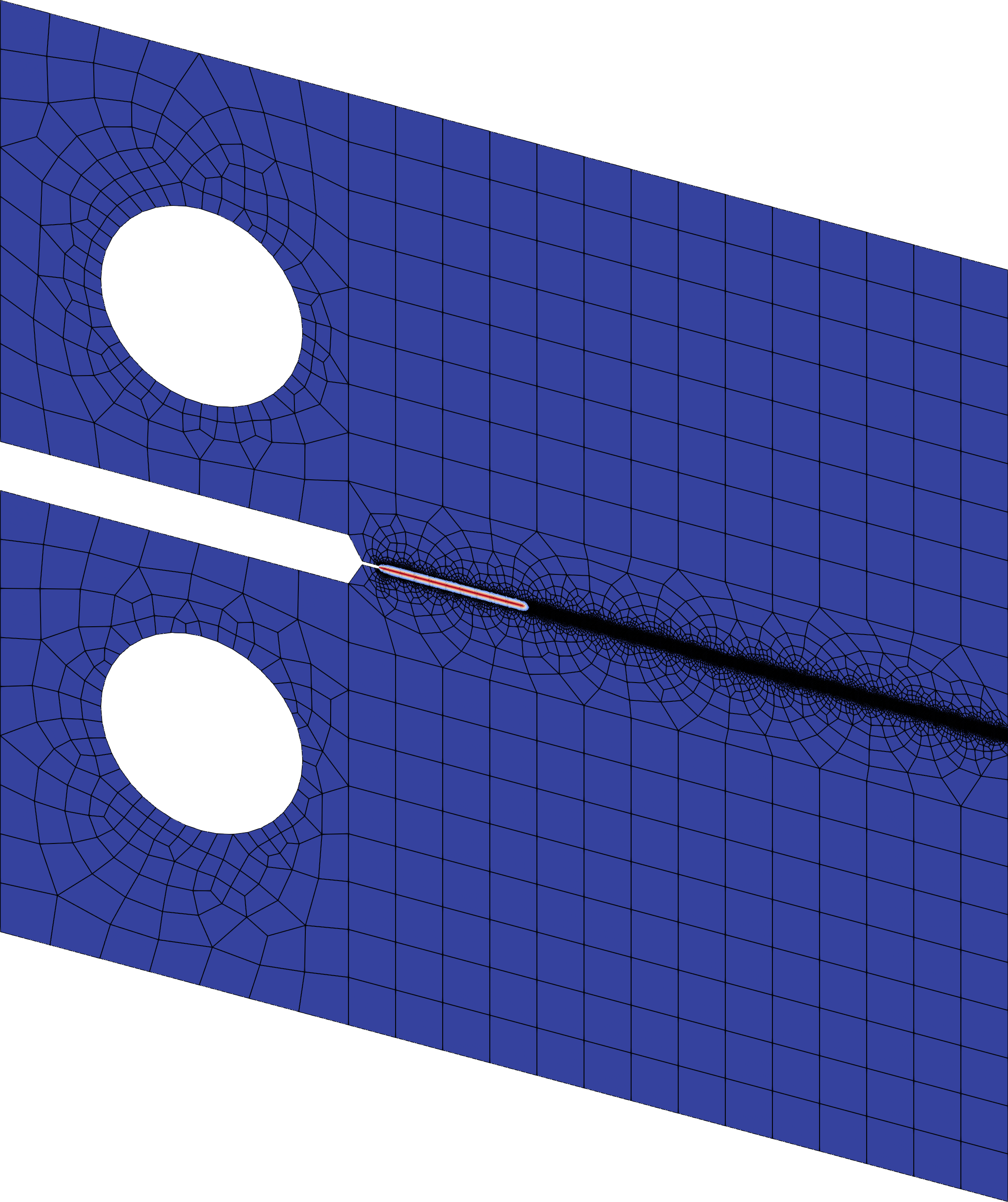}};
            \node[anchor=center] at (2.4,0) {\includegraphics{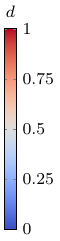}};
        \end{tikzpicture}
        \label{fig:phasefield_ct}
    }
    \caption{Geometry, boundary conditions (a) and finite element mesh showing the phase-field contour at an exemplary load step during fatigue crack propagation (b) for the CT test.}
    \label{fig:ct_test}
\end{figure}
We consider a compact tension (CT) specimen with geometry and dimensions in Fig. \ref{fig:specimen_ct}, assuming plane-stress conditions (however without explicitly enforcing $\sigma_{33}=0$ as the phase field evolves, see \cite{Li_phase_2021} for more details).
The test is performed in load control applying the boundary conditions in Fig. \ref{fig:phasefield_ct}. Following the parameters in Fig. \ref{fig:loading_curve_discretization}, the load cycles have $P_{\min}=0$~N and $P_{\max}$ ranging between $50$~N and $150$~N to cover different fatigue regimes.
 We adopt the parameters in \cite{carrara_framework_2020}, namely Young's modulus $E=6000$~MPa, Poisson's ratio $\nu=0.22$, regularizing length $\ell=0.2$~mm, and fracture toughness $\mathcal{G}_{\text{c}} = 2.28$~MPa~mm leading to a critical stress intensity factor $K_{\text{Ic}}=3.69$~MPa~$\sqrt{\text{m}}$.
Also, we compare the results for both the AT1 and AT2 local dissipation functions, we choose the spectral energy decomposition  \cite{miehe_thermodynamically_2010} and we set $\lambda_{\text{II}} = \lambda_{\text{III}}$ if not stated otherwise.
For the fatigue degradation function (\ref{eq:fatigue_deg_fun_generalized}) we set $\bar{\alpha}_{\text{th}} = 60$~N~mm$^{-2}$ and $p=2$.
The mesh with $20736$ bilinear quadrilateral elements and $20844$ nodes is locally refined in the area of the expected crack path with $\ell/h \approx 5$, as depicted in Fig. \ref{fig:phasefield_ct}.

\subsubsection{Speedup and accuracy}
\begin{figure*}
    \centering
    \subfloat[]{
        \includegraphics{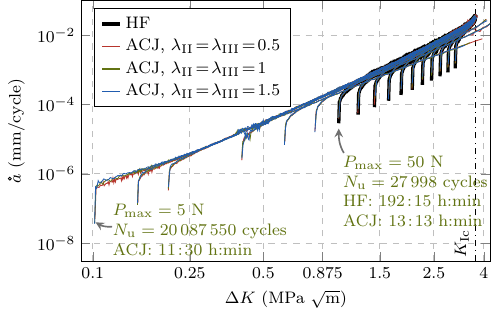}
        \label{fig:numerical_results_ct_cgr_AT1}
    }
    \subfloat[]{
        \includegraphics{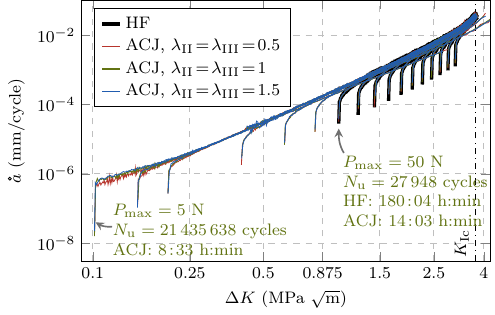}
        \label{fig:numerical_results_ct_cgr_AT2}
    }
    \caption{Comparison of crack growth rate curves for the CT test with the AT1 (a) and AT2 (b) model, as obtained from HF computations and the proposed ACJ scheme with different speedup factors. The HF-computed curves are limited to the load levels for which failure is reached  within a computational time of two weeks (i.e., $P_{\max} = 50, 60, 70, 80, 90, 100, 110, 120, 130, 140, 150$~N). Additionally, the ACJ approach allows to compute the crack growth rate curve for the load levels $P_{\max} = 5, 7.5, 10, 15, 20, 30, 40$~N.}
    \label{fig:numerical_results_ct_cgr}
\end{figure*}
In Fig. \ref{fig:numerical_results_ct_cgr}, the predicted crack growth rates are presented at the load levels $P_{\max} = 50$ to $150$~N for the CT test, obtained with HF and accelerated computations.
For the ACJ, we investigate also the role of the speedup parameters by comparing the results obtained with $\lambda_{\text{II}}\!=\!\lambda_{\text{III}} = 0.5, 1, 1.5$.
We can observe an almost perfect overlap of the crack growth rate curve independently of the load applied or of the used speedup parameters $\lambda_{\text{II,III}}$.
In particular, the ACJ is able to correctly reproduce the initial crack nucleation phase, the stable crack propagation (i.e., the Paris regime) and the final unstable failure stage (Fig. \ref{fig:numerical_results_ct_cgr}).
The load level $P_{\max} = 50$~N is the lowest one attainable with a HF computation within a reasonable amount of time, i.e., in about $8$~days ($7.5$~days), yielding a fatigue life of $N_{\text{u}} = 27998$~cycles ($27948$~cycles) for the AT1 (AT2) model, which can still be considered as LCF regime.
In contrast, the ACJ computation at the same load level takes about $13$~hours ($14$~hours) for $\lambda_{\text{II}} = \lambda_{\text{III}} = 1$ and the AT1 (AT2) model, cutting the CPU time by $93$~\% ($92$~\%) with of a deviation in terms of fatigue life (i.e., the total number of cycles until failure) of only $1.1$~\% ($0.3$~\%).
For the ACJ with $\lambda_{\text{II}} = \lambda_{\text{III}} = 1$, the number of resolved cycles is $N_{\text{r}} = 1045$~cycles, that is $3.7$~\% ($1053$~cycles, $3.8$~\%) of the total number of cycles.
Similar considerations apply to the other load levels (Tab. \ref{tab:CT_compar_LCF}), demonstrating that the ACJ approach is able to provide fast, yet accurate results.
Tab. \ref{tab:CT_compar_LCF} reports CPU time, number of resolved cycles $N_{\text{r}}$ and estimated fatigue life $N_{\text{u}}$, only for the AT1 model since the results for the AT2 model are very similar.

\begin{table*}
    \caption{Predicted fatigue life $N_{\text{u}}$, number of resolved cycles $N_{\text{r}}$ and CPU time for the CT test comparing HF computations with the ACJ for different speedup parameters in the LCF regime (AT1 model).}
    \label{tab:CT_compar_LCF}
    \centering
    \footnotesize
    \begin{tabularx}{\textwidth}{l|ll|lll|lll|lll}
        \toprule
        \multirow{2}{*}{$P_{\max}$}
            &\multicolumn{2}{c}{HF}
            &\multicolumn{3}{c}{ACJ, $\lambda_{\text{II}}\!=\!\lambda_{\text{III}}\!=\!0.5$}
            &\multicolumn{3}{c}{ACJ, $\lambda_{\text{II}}\!=\!\lambda_{\text{III}}\!=\!1$}
            &\multicolumn{3}{c}{ACJ, $\lambda_{\text{II}}\!=\!\lambda_{\text{III}}\!=\!1.5$}\\
            &$N_{\text{u}}$ &time (h:min) &$N_{\text{u}}$ &$N_{\text{r}}$ &time (h:min) &$N_{\text{u}}$ &$N_{\text{r}}$ &time (h:min) &$N_{\text{u}}$ &$N_{\text{r}}$ &time (h:min)\\
        \midrule
        150	&403        &8:17       &404            &228        &5:35       &408        &157            &3:26       &415        &125        &2:39      \\
            &(100\%)    &(100\%)    &(+0.3\%)       &(56\%)     &(67\%)     &(+1\%)     &(38\%)         &(42\%)     &(+3\%)  &(30\%)  &(32\%) \\
        140	&600        &11:26      &604            &318        &7:17       &606        &209            &4:23       &615        &160        &3:13      \\
            &(100\%)    &(100\%)    &(+0.7\%)       &(53\%)     &(64\%)     &(+1\%)     &(34\%)         &(38\%)     &(+3\%)   &(26\%)  &(28\%) \\
        130	&867        &14:48      &870            &417        &8:34       &874        &272            &5:22       &887        &206        &4:06       \\
            &(100\%)    &(100\%)    &(+0.4\%)       &(48\%)     &(58\%)     &(+0.8\%)  &(31\%)          &(36\%)     &(+2\%)  &(23\%)  &(28\%)  \\
        120	&1237       &20:36      &1241           &528        &9:52       &1248       &339            &6:21       &1269       &252        &4:43      \\
            &(100\%)    &(100\%)    &(+0.3\%)       &(43\%)     &(48\%)     &(+0.9\%)  &(27\%)          &(31\%)     &(+3\%)  &(20\%)  &(23\%) \\
        110	&1781       &25:34      &1788           &671        &13:09       &1804       &421            &8:26       &1828       &314        &6:22      \\
            &(100\%)    &(100\%)    &(+0.4\%)       &(38\%)     &(51\%)     &(+1\%)     &(23\%)         &(33\%)     &(+3\%)  &(17\%)  &(25\%) \\
        100	&2582       &35:26      &2589           &826        &15:15      &2601       &511            &9:27       &2629       &370        &7:22      \\
            &(100\%)    &(100\%)    &(+0.3\%)       &(32\%)     &(43\%)     &(+0.7\%)  &(20\%)          &(27\%)     &(+2\%)  &(14\%)  &(21\%) \\
         90&3815       &47:17      &3824            &1003       &16:17      &3860       &602            &10:12       &3888       &433        &7:54      \\
            &(100\%)    &(100\%)    &(+0.2\%)       &(26\%)     &(34\%)     &(+1\%)     &(16\%)         &(22\%)     &(+2\%)  &(11\%)  &(17\%) \\
         80 &5801       &67:14      &5815           &1199       &17:53      &5875       &701            &11:23      &5914       &499        &8:05       \\
            &(100\%)    &(100\%)    &(+0.2\%)       &(21\%)     &(27\%)     &(+1\%)     &(12\%)         &(17\%)     &(+2\%)  &(8\%)   &(12\%) \\
         70 &9187       &97:44      &9212           &1417       &19:43      &9256       &805            &11:45      &9358       &565        &8:22      \\
            &(100\%)    &(100\%)    &(+0.3\%)       &(15\%)     &(20\%)     &(+0.8\%)  &(9\%)           &(12\%)     &(+2\%)  &(6\%)   &(9\%)  \\
         60 &15405      &121:35     &15450          &1650       &21:21      &15558      &921            &15:20      &15726      &637        &9:06       \\
            &(100\%)    &(100\%)    &(+0.3\%)       &(11\%)     &(18\%)     &(+1\%)     &(6\%)          &(13\%)     &(+2\%)  &(4\%)   &(7\%)  \\
         50 &27998      &192:15     &28084          &1912       &22:38      &28293      &1045           &13:13      &28551      &713     &9:31      \\
            &(100\%)    &(100\%)    &(+0.3\%)       &(7\%)      &(12\%)     &(+1\%)     &(4\%)          &(7\%)      &(+2\%)  &(3\%)    &(5\%)  \\
        \bottomrule
    \end{tabularx}
\end{table*}

The proposed ACJ scheme allows to extend the study to HCF, whose computational cost is prohibitive for a HF computation.
Fig. \ref{fig:numerical_results_ct_cgr} shows the predicted crack growth rates for additional load levels down to $P_{\max} = 5$~N, for which a predicted fatigue life of around $2\cdot10^7$~cycles is obtained within less than a day.
The small change of computational time when comparing the case $P_{\max} = 5$~N with $P_{\max} = 50$~N is due to the slow evolution rate of the system, which allows for larger jumps.
For such long fatigue lives, some minor oscillations of the crack growth rate curve can be observed in the early stable crack propagation stage.
For the lowest load levels in the VHCF regime, no distinctive unstable crack propagation branch of the crack growth rate curve can be observed anymore, and the branch of stable crack propagation (i.e., the Paris regime) appears to be slightly less steep.
The details of the computations for $P_{\max} = 5, 7.5, 10, 15, 20, 30, 40$ of Fig. \ref{fig:numerical_results_ct_cgr} are summarized in Tab. \ref{tab:CT_compar_HCF}, again only for the AT1 model.
We can observe that the computational time barely depends on the load level, while it is influenced by the speedup factors, and likewise for the number of resolved cycles, which is in line with the discussion in Section \ref{sec:summary_and_estimation_of_N_r}.
For the lowest load level $P_{\max} = 5$ with $\lambda_{\text{II,III}}=1.5$, only $1006$ of the total $N_{\text{u}}=19\,565\,228$~cycles are resolved ($0.01$~\%) within only around $6$~hours of computational time. It can be concluded that, in these fatigue regimes, the proposed acceleration scheme enables computations which are otherwise not feasible.

\begin{table*}
    \caption{Predicted fatigue life $N_{\text{u}}$, number of resolved cycles $N_{\text{r}}$ and CPU time for the CT test obtained with the ACJ for different speedup parameters in the  HCF regime with no reference results from HF computations available (AT1 model).}
    \label{tab:CT_compar_HCF}
    \centering
    \footnotesize
    \begin{tabularx}{\textwidth}{l|XlX|XlX|XlX}
        \toprule
        \multirow{2}{*}{$P_{\max}$}
            &\multicolumn{3}{c}{ACJ, $\lambda_{\text{II}}\!=\!\lambda_{\text{III}}\!=\!0.5$}
            &\multicolumn{3}{c}{ACJ, $\lambda_{\text{II}}\!=\!\lambda_{\text{III}}\!=\!1$}
            &\multicolumn{3}{c}{ACJ, $\lambda_{\text{II}}\!=\!\lambda_{\text{III}}\!=\!1.5$}\\
            &$N_{\text{u}}$ &$N_{\text{r}}$ &time (h:min) &$N_{\text{u}}$ &$N_{\text{r}}$ &time (h:min) &$N_{\text{u}}$ &$N_{\text{r}}$ &time (h:min)\\
        \midrule
        40      &57513      &2197   &24:20  &57836      &1171   &13:47   &58511     &793   &9:30  \\
        30      &142453     &2504   &24:13  &142845     &1305   &13:40   &145023    &881   &9:36  \\
        20      &497044     &2837   &23:18  &499065     &1459   &12:28   &510366    &973   &8:36  \\
        15      &1207717    &3025   &22:39  &1208843    &1536   &12:27   &1219572   &1029  &9:12  \\
        10      &3859061    &3189   &20:12  &3781558    &1602   &10:32   &3776734   &1058  &8:08  \\
        7.5     &8161776    &3200   &24:07  &7835092    &1592   &10:25   &7640258   &1054  &6:57  \\
         5      &21595566   &3114   &18:47  &20087550   &1509   &11:30   &19565228  &1006  &6:06  \\
        \bottomrule
    \end{tabularx}
\end{table*}

\subsubsection{Comparison with other approaches}
We now compare our ACJ scheme with other available approaches in terms of accuracy and speedup.
In particular, we consider two alternatives to determine the cycle-jump size $\Delta N$. 
In particular, we consider two alternatives to determine the cycle-jump size $\Delta N$. 
In particular, we consider two alternatives to determine the cycle-jump size $\Delta N$.
Firstly, we use a fixed cycle jump (FCJ), where a jump with $\Delta N = \text{const.}$ is performed after each block of resolved cycles, where we choose sizes of $\Delta N = 100, 1000, 10000$~cycles.
Secondly, we use the cycle jump based on a criterion on the extrapolation of $\bar{\alpha}$ proposed in \cite{seles_general_2021}, referred to as ECJ in the following, which determines the cycle-jump size as
\begin{equation}
    \begin{aligned}
        &\Delta N (N) = \min \biggl\{\\
        &q \cdot \underset{\bs{x} \in \Omega}{\min}\left\{ \frac{\vert \Delta \bar{\alpha} (\bs{x},N) \vert}{\vert \Delta \bar{\alpha} (\bs{x},N) - \Delta \bar{\alpha} (\bs{x},N\!-\!1) \vert} \right\}, \Delta N_{\max} \biggr\}\\[1em]
        &\quad\text{with}\quad\Delta \bar{\alpha} (\bs{x},N) = \vert \bar{\alpha} (\bs{x},N) - \bar{\alpha} (\bs{x},N\!-\!1) \vert \text{ .}
    \end{aligned}
\end{equation}
Here $q$ is a user-defined parameter governing the maximum relative deviation between the rate of the extrapolated quantity prior to and after the jump.
As local variable to be extrapolated we use $\bar{\alpha}(\bs{x},N)$ as well, and for a fair comparison, the extrapolation is performed using (\ref{eq:FD_extrapolation}) and a set of $N_{\text{s}}$ resolved cycles.
Since in \cite{seles_general_2021} no value for the user parameter is indicated, here $q = 1$ is used.
Further, we added an upper limit for the jump extension $\Delta N_{\max}$ to avoid too optimistic cycle jumps and to increase the stability of the scheme in e.g. the very beginning or in the transition between the different stages of the computations.
Since there is no possibility to determine \textit{a priori} a feasible limit, we use $\Delta N_{\max} = 1000, 10000$~cycles.
To allow for a fair comparison, all the approaches are implemented within the same code, while the analyses are performed on comparable hardware.
Only the computations which reach a stress intensity factor amplitude of at least $70$~\% of $K_{\text{Ic}}$ are considered in the comparison;
the runs during which the computations stop prior to that point due to numerical issues are not considered as valid.

\begin{figure}[t]
    \centering
    \hspace*{1cm}\includegraphics{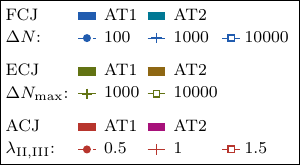}\\[0.5em]
    \subfloat[]{
        \includegraphics{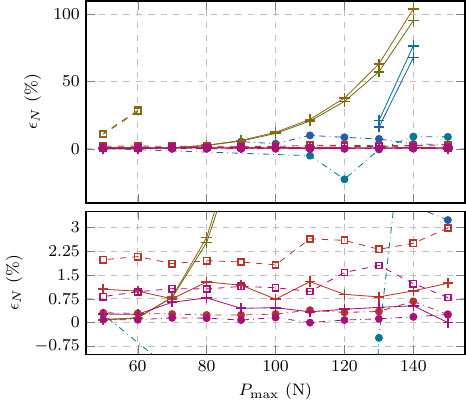}
        \label{fig:numerical_results_ct_epsilon_N}
        }\\
    \subfloat[]{
        \includegraphics{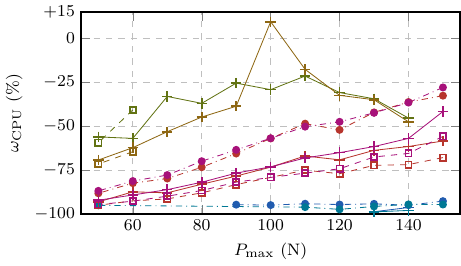}
        \label{fig:numerical_results_ct_omega_CPU}
    }
    \caption{Comparison of the FCJ, ECJ and ACJ approaches to HF computations in terms of accuracy $\epsilon_N$ (a) and speedup $\omega_{\text{CPU}}$ (b) for the CT test in the LCF regime.}
    \label{fig:numerical_results_ct_performance_LCF}
\end{figure}

To provide a more insightful evaluation of the performance of the approaches, we introduce first the fatigue life prediction error $\epsilon_N$ as
\begin{equation}
    \epsilon_N := \frac{N_{\text{u}}}{N_{\text{u,HF}}}-1
\end{equation}
to compare the accuracy of the accelerated computations to the HF computations wherever possible.
Secondly, the speedup is quantified in terms of CPU time $\omega_{\text{CPU}}$ as
\begin{equation}
    \omega_{\text{CPU}} := \frac{\text{CPU time}}{\text{CPU time}_{\text{HF}}} - 1 \text{ ,}
\end{equation}
and finally, since for peak loads $P_{\max} < 50$~N, no HF computation is available, the speedup is obtained in terms of resolved cycles $N_{\text{r}}$ for the estimated fatigue life $N_{\text{u}}$ as
\begin{equation}
    \omega_{N} = \frac{N_{\text{u}}}{N_{\text{r}}} \text{ .}
\end{equation}
For the ACJ, the FCJ, and the ECJ approach, the accuracy in terms of $\epsilon_N$ and the speedup in terms of $\omega_{\text{CPU}}$ in comparison to the HF computations are plotted in Figs. \ref{fig:numerical_results_ct_epsilon_N} and \ref{fig:numerical_results_ct_omega_CPU}, respectively.

First, we note that some data points of the FCJ and ECJ approach are missing due to the impossibility to reach convergence and hence $70$~\% of $K_{\text{Ic}}$ after a too optimistic cycle jump, either within the Newton-Raphson iterative solution of the equilibrium or damage equation, or within the staggered scheme within the allotted maximum number of iterations ($250$ for each).
This is the case for all the FCJ computations with $\Delta N=10000$ and almost all with $\Delta N = 1000$~cycles, highlighting the lack of robustness of this approach and the critical role played by the choice of $\Delta N$.
$\Delta N = 100$~cycles seems to be the most suitable choice for an LCF regime, while the obtained speedup in terms $\omega_{\text{CPU}}$ is limited (Fig. \ref{fig:numerical_results_ct_omega_CPU}), and the error $\epsilon_N$ is quite high, reaching up to about $77$~\% (Fig. \ref{fig:numerical_results_ct_epsilon_N}).
Considering the ECJ, none of the tested $\Delta N_{\max}$ values ensures convergence for all test cases, highlighting the key role of this parameter for which no estimate is available.
In particular, when higher values of $\Delta N_{\max}$ are selected, convergence issues are observed in the LCF regimes along with higher fatigue life estimation errors $\epsilon_N$ (Fig. \ref{fig:numerical_results_ct_epsilon_N}), while for low $\Delta N_{\max}$ values, a small speedup is obtained in the HCF regime.
Thus, these approaches achieve either a good speedup or a good accuracy with an appropriate estimate for their parameters, but not both, while numerical issues are encountered with a bad estimate for their parameters.
Conversely, the proposed approach reaches convergence for any of the tested speedup parameters $\lambda_{\text{II,III}}$ while keeping the error $\epsilon_N$ consistently below around $3$~\%, clearly depending on the choice of $\lambda_{\text{II,III}}$.
Moreover, the obtained speedup is always higher and growing at an increasing rate towards the HCF regime.
Comparing Fig. \ref{fig:numerical_results_ct_epsilon_N} ($\epsilon_N$) with Fig. \ref{fig:numerical_results_ct_omega_CPU} ($\omega_{CPU}$) offers a further confirmation of the role of $\lambda_{\text{II,III}}$ as parameters governing the trade-off between accuracy and efficiency.

\begin{figure}
    \begin{center}
        \includegraphics{numerical_results_ct_legend_comparison.pdf}
    \end{center}
    \vspace*{0.1cm}
    \subfloat[]{
        \includegraphics{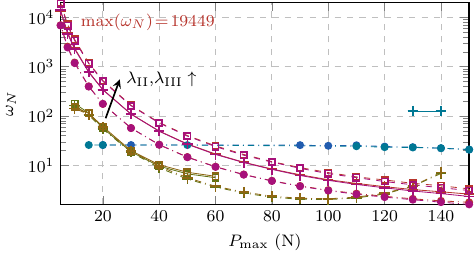}
        \label{fig:numerical_results_ct_omega_N_P}
    }\\
    \subfloat[]{
        \includegraphics{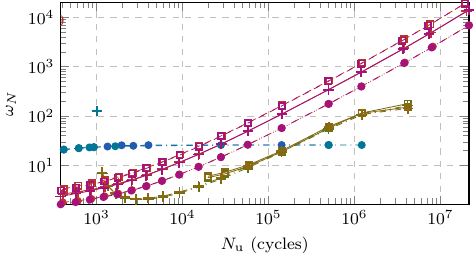}
        \label{fig:numerical_results_ct_omega_N_N}
    }
    \caption{Comparison between the FCJ, ECJ and ACJ approaches in terms of speedup $\omega_N$ for different load levels (a) and number of cycles until failure (b) for the CT test.}
    \label{fig:numerical_results_ct_omega_N}
\end{figure}

Extending the performance analysis to the HCF regime with the additional load levels by evaluating $\omega_N$ in Fig. \ref{fig:numerical_results_ct_omega_N}, the same conclusions can be drawn.
In particular, Fig. \ref{fig:numerical_results_ct_omega_N_P} shows $\omega_N$ for different load levels, while Fig. \ref{fig:numerical_results_ct_omega_N_N} gives $\omega_N$ as a function of the total fatigue life.
As observed before, some data points are missing for the FCJ and ECJ either due to the mentioned numerical issues, or since the speedup obtained was too small such that the computation did not reach the final stage within a week of computational time.
In contrast, the proposed ACJ consistently reaches the final stage of the fatigue life even for the lowest load levels, showing a robust behavior in dependency of the speedup factors.
Also, the value of $\omega_N$ increases for decreasing $P_{\max}$ until reaching, for $\lambda_{\text{II,III}} = 1$, a speedup of more than four orders of magnitude, namely $\omega_N = 1.3 \cdot 10^4$ ($1.4 \cdot 10^4$) for the AT1 (AT2) model.
This means that on average, one cycle every $13312$ ($14038$) cycles needs to be resolved.
Similar observations can be made for the analysis with  $\lambda_{\text{II,III}} < 1$ ($\lambda_{\text{II,III}} > 1$) for which a lower (higher) speedup is obtained in spite of the negligible variation of the crack growth rate curve (Fig. \ref{fig:numerical_results_ct_cgr}) and fatigue life (Tab. \ref{tab:CT_compar_HCF}).
For speedup factors of $\lambda_{\text{II,III}} = 1.5$ even higher values up to almost $\omega_N = 2 \cdot 10^4$ are obtained.

\subsubsection{Behavior in different stages and fatigue life regimes}
\begin{figure}
    \centering
    \includegraphics{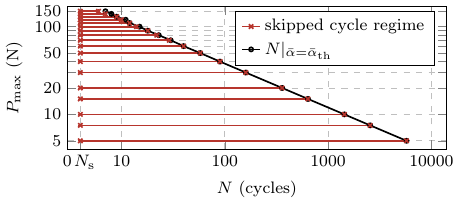}
    \caption{Stage I cycle-jump size for the CT test and comparison with the stage extension in terms of number of cycles obtained adopting the AT1 model.}
    \label{fig:numerical_results_ct_stage_I}
\end{figure}
Next, we analyze the behavior of the proposed ACJ scheme within the different stages of the fatigue life.
Fig. \ref{fig:numerical_results_ct_stage_I} illustrates for stage I the extension of the cycle jump as a function of the maximum applied load $P_{\max}$ for the AT1 model, since for the AT2 model, the same conclusions can be drawn.
The black data points indicate the cycle at which the fatigue threshold $\bar{\alpha}_{\text{th}}$ is met for the first time $N|_{\bar{\alpha}=\bar{\alpha}_{\text{th}}}$.
The first data point at $N_{\text{s}}$ indicates the start of the cycle jump while the second data point represents the cycle to which the algorithm jumps, which is either slightly below $N|_{\bar{\alpha}=\bar{\alpha}_{\text{th}}}$ or matching it exactly.
At high load levels ($P_{\max} = 150$~N), this allows for a  small jump of only $2$~cycles after the first $N_{\text{s}}$ resolved cycles, while the cycle-jump size significantly increases as the applied load decreases.
This applies to both AT1 and AT2.

\begin{figure*}
    \centering
    \subfloat[]{
        \includegraphics{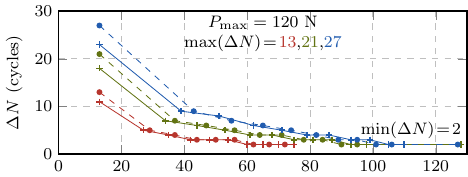}
        \label{fig:numerical_results_ct_DN_II_120}
    }\hfill
    \subfloat[]{
        \includegraphics{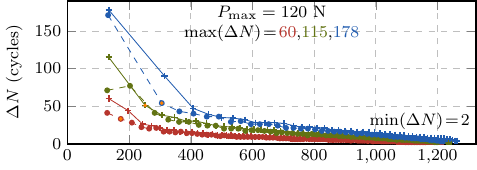}
        \label{fig:numerical_results_ct_DN_III_120}
    }
    \newline
    \subfloat[]{
        \includegraphics{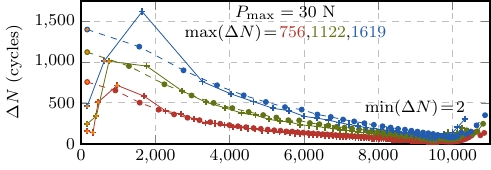}
        \label{fig:numerical_results_ct_DN_II_30}
    }
    \subfloat[]{
        \includegraphics{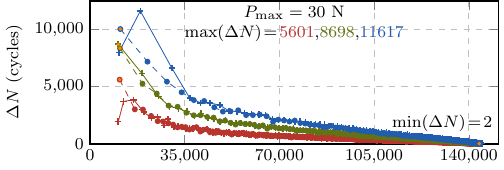}
        \label{fig:numerical_results_ct_DN_III_30}
    }
    \newline
    \subfloat[]{
        \includegraphics{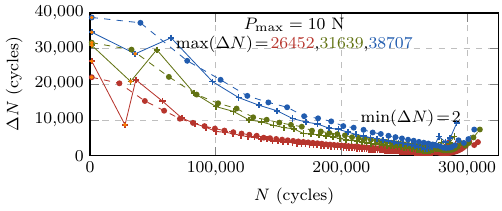}
        \label{fig:numerical_results_ct_DN_II_10}
    }
    \subfloat[]{
        \includegraphics{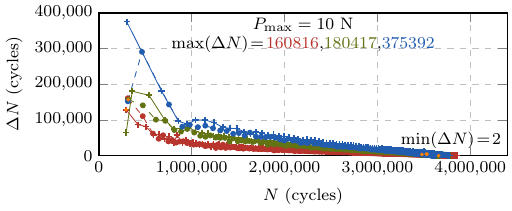}
        \label{fig:numerical_results_ct_DN_III_10}
    }
    \\[-2em]
    \includegraphics{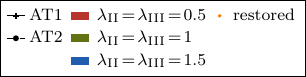}
    \caption{Size of the performed cycle jumps throughout the fatigue life in stage II (a,c,e) and in stage III (b,d,f) at selected load levels $P_{\max} = 10, 30, 120$~N for the CT test. The points are connected by lines for better visibility.}
    \label{fig:numerical_results_ct_DN}
\end{figure*}

\begin{figure*}
    \centering
    \subfloat[]{
        \includegraphics{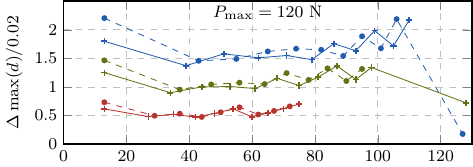}
        \label{fig:numerical_results_ct_DLambda_II_120}
    }
    \subfloat[]{
        \includegraphics{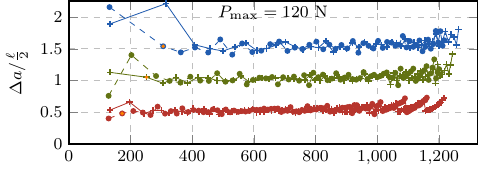}
        \label{fig:numerical_results_ct_DLambda_III_120}
    }
    \newline
    \subfloat[]{
        \includegraphics{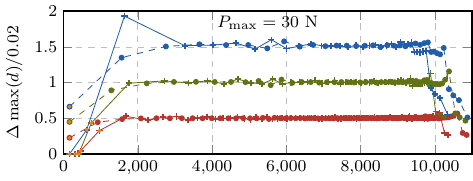}
        \label{fig:numerical_results_ct_DLambda_II_30}
    }
    \subfloat[]{
        \includegraphics{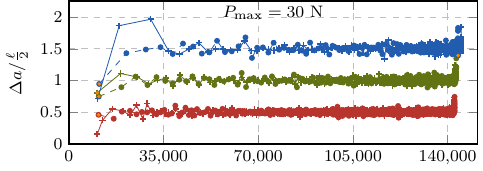}
        \label{fig:numerical_results_ct_DLambda_III_30}
    }
    \newline
    \subfloat[]{
        \includegraphics{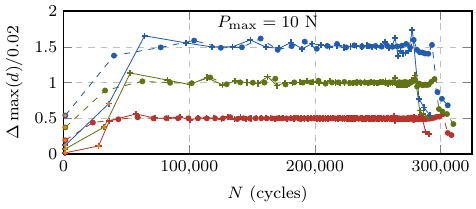}
        \label{fig:numerical_results_ct_DLambda_II_10}
    }
    \subfloat[]{
        \includegraphics{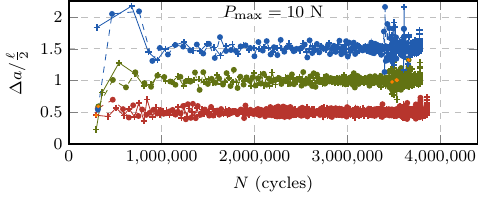}
        \label{fig:numerical_results_ct_DLambda_III_10}
    }
    \\[-2em]
    \includegraphics{numerical_results_ct_legend_stagewise_restored.pdf}
    \caption{Obtained increments of $\Lambda$ during the performed cycle jumps in stage II (a,c,e) and in stage III (b,d,f) at selected load levels $P_{\max} = 10, 30, 120 $~N for the CT test. The points are connected by lines for better visibility.}
    \label{fig:numerical_results_DLambda_II_III}
\end{figure*}

For the analysis of stage II and stage III, we select three load levels representative of different fatigue regimes, namely the LCF regime at $P_{\max} = 120$~N, an intermediate regime at $P_{\max} = 30$~N, and the HCF regime at $P_{\max} = 10$~N.
Figs. \ref{fig:numerical_results_ct_DN} (a,c,e) and \ref{fig:numerical_results_ct_DN} (b,d,f) show the performed cycle jumps $\Delta N$ at these load levels during stage II and stage III, respectively.
The proposed algorithm performs well for these different regimes characterized by a fatigue life $N_{\text{u}}$ ranging from $10^3$~cycles to more than $10^6$~cycles.
It is evident that the extension of the jumps is larger at the beginning of both stage II and III due to the lower evolution rate of the system and reaches up to more than $10^2$, $10^4$ and $10^5$ cycles respectively for $P_{\max} = 120$~N, $30$~N and $10$~N.
The amount of jumped cycles then gradually decreases as the rate of change of the global control quantity $\Lambda$ increases, until reaching the minimum jump size $\Delta N = 2$ at the end of each stage.
This is the case even in the HCF regime, where the largest cycle jump is $375392$~cycles.
As demonstrated by Figs. \ref{fig:numerical_results_DLambda_II_III} (a-f), this change of the jump size is needed to ensure that the variation of the global control variable during a jump $\Delta \Lambda$ remains close to the imposed value $\Delta \bar{\Lambda}$, highlighting the reliability of the proposed cycle-jump criterion.
Note that at the very end of stages II and III, the system evolves so quickly that the cycle-jump criterion would give $\Delta N < 2$, therefore the algorithm reverts back to a HF computation as detailed in Section \ref{sec:automatic_return_to_hf_computation}.
For the case illustrated here, rejecting a cycle jump which encountered convergence issues or gave a too large $\Delta \Lambda$ during the trial cycle (Section \ref{sec:algorithmic_aspects}) is limited to either the beginning or the end of a stage, as indicated with the orange data points in Figs. \ref{fig:numerical_results_ct_DN} and \ref{fig:numerical_results_DLambda_II_III}.

Fig. \ref{fig:numerical_results_DLambda_II_III} clarifies also the role of the speedup parameters $\lambda_{\text{II}}$ and $\lambda_{\text{III}}$, which modulate the target value $\Delta \bar{\Lambda}$ and hence, reduce or increase the jump size. An increase of $\lambda_{\text{II,III}}>1$ leads to less computational effort, but too optimistic cycle jumps needing correction are more frequent.
Based on the performed computations, we suggest to keep $\lambda_{\text{II,III}}\leq 1.5$ to preserve the robustness and accuracy of the results.
Also, the proposed ACJ algorithm behaves similarly for both AT1 and AT2 dissipation functions without modifying the target values $\Delta \bar{\Lambda}$, and for any investigated value of $\lambda_{\text{II}}$ and $\lambda_{\text{III}}$, further confirming the robustness of the method.

Tab. \ref{tab:cyclejump_behavior_details} gives a quantitative comparison of the performed cycle jumps in the different stages, only for the AT1 model since the results for the AT2 model are very similar.
Clearly, most of the CPU time is spent in stage III, while stage I takes the smallest time.

\begin{table*}
    \caption{Comparison of the different stages in terms of cycle-jump sizes, number of resolved cycles and percentage of total CPU time spent within each of the three stage for the CT test (AT1 model).}
    \label{tab:cyclejump_behavior_details}
    \centering
    \footnotesize
    \begin{tabularx}{\textwidth}{l|l|l|Xll|Xll|Xll}
        \toprule
        \multirow{2}{*}{$P_{\text{max}}$}
            &\multirow{2}{*}{$\lambda_{\text{II,III}}$}
            &\multirow{2}{*}{$N_{\text{u}}$}
            &\multicolumn{3}{c}{stage I}
            &\multicolumn{3}{c}{stage II}
            &\multicolumn{3}{c}{stage III}\\
        &&&$\Delta N$ &$N_{\text{r,I}}$ &time share (\%) &\# jumps &$N_{\text{r,II}}$ &time share (\%) &\# jumps &$N_{\text{r,III}}$ &time share (\%)\\
        \midrule
        \multirow{3}{*}{$10$}
        &0.5    &3859061    &1434   &5   &0.03\%   &101 &403    &3\%   &693  &2781   &97\%\\
        &1      &3781558    &1434   &5   &0.06\%   &52  &207    &3\%   &347  &1390   &97\%\\
        &1.5    &3776734    &1434   &5   &0.08\%   &36  &143    &3\%   &226  &910    &97\%\\
        \multirow{3}{*}{$30$}
        &0.5    &142453     &156    &5   &0.03\%   &98  &391    &3\%   &522  &2108   &97\%\\
        &1      &142845     &156    &5   &0.05\%   &52  &207    &4\%   &272  &1093   &96\%\\
        &1.5    &145023     &156    &5   &0.07\%   &36  &143    &3\%   &182  &733    &97\%\\
        \multirow{3}{*}{$120$}
        &0.5    &1241       &6      &6   &0.1\%   &10  &95     &4\%   &102   &427    &96\%\\
        &1      &1248       &6      &6   &0.2\%   &12  &75     &6\%   &65    &258    &93\%\\
        &1.5    &1269       &6      &6   &0.3\%   &11  &64     &8\%    &46   &182     &92\%\\
        \bottomrule
    \end{tabularx}
\end{table*}

\begin{figure}
    \centering
    \includegraphics{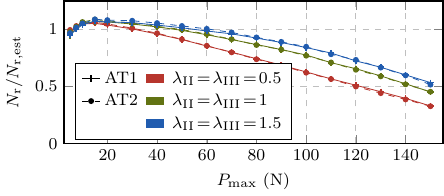}
    \caption{Comparison between the actual number of resolved cycles $N_{\text{r}}$ and the estimate obtained using (\ref{eq:nresolved_estimation}) for the CT test.}
    \label{fig:numerical_results_ct_N_r}
\end{figure}

In Fig. \ref{fig:numerical_results_ct_N_r} we compare the estimate for the number of cycles which need to be computed $N_{\text{r,est}}$ (Section \ref{sec:summary_and_estimation_of_N_r}) with the number of actually resolved cycles $N_{\text{r}}$.
At the higher load levels, the $N_{\text{r}}/N_{\text{r,est}}$ ratio is around $0.5$, meaning that only about half of the estimated number of HF cycles are actually computed.
This difference is due to the fact that, in the LCF regime, the system experiences a non-negligible evolution also during the HF computed cycles, which is neglected in \eqref{eq:nresolved_estimation}.
The higher the fatigue life, the more the ratio $N_{\text{r}}/N_{\text{r,est}}$ tends towards one, which means that the algorithm is performing as expected, due to the obtained increments $\Delta \bar{\Lambda}$ being close to the target value.

\begin{figure*}
    \centering
    \subfloat[]{
        \hspace{0.125cm}
        \includegraphics{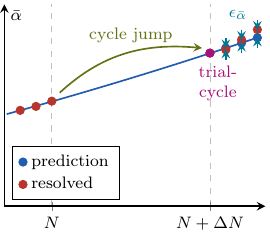}
        \hspace{0.125cm}
        \label{fig:prediction_error}
    }
    \subfloat[]{
        \begin{tikzpicture}[every node/.style={inner sep=0pt,outer sep=0pt}]
            \footnotesize
            \begin{scope}[scale=0.75]
                \node[anchor=center] at (0,0) {\includegraphics[width=4.5cm]{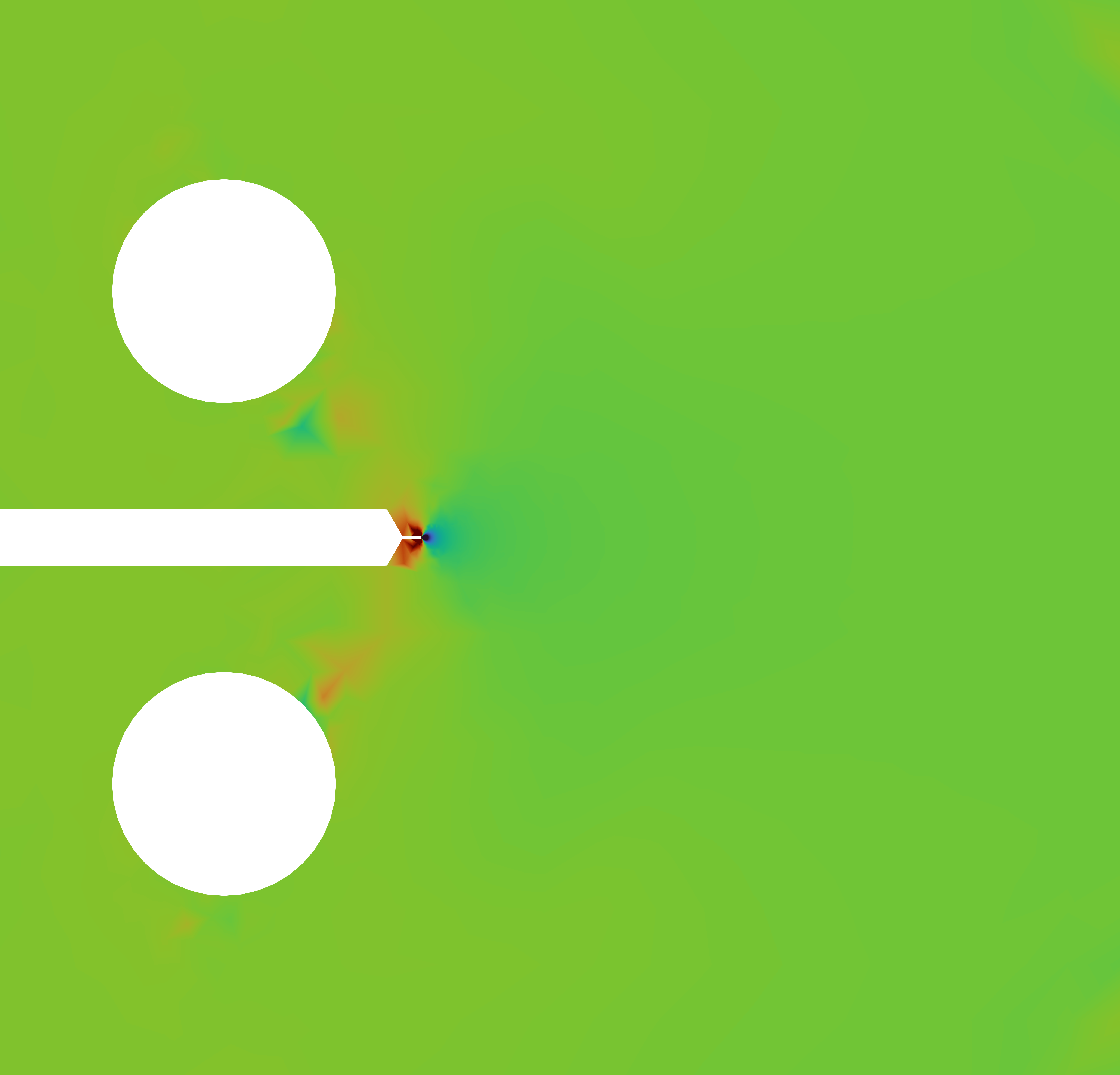}}; 
                \draw[black] (-0.825,-0.1) rectangle (-0.6,0.1);
                \node[fill=white, anchor=north east] at (2.75,2.63) {\includegraphics[width=2.25cm]{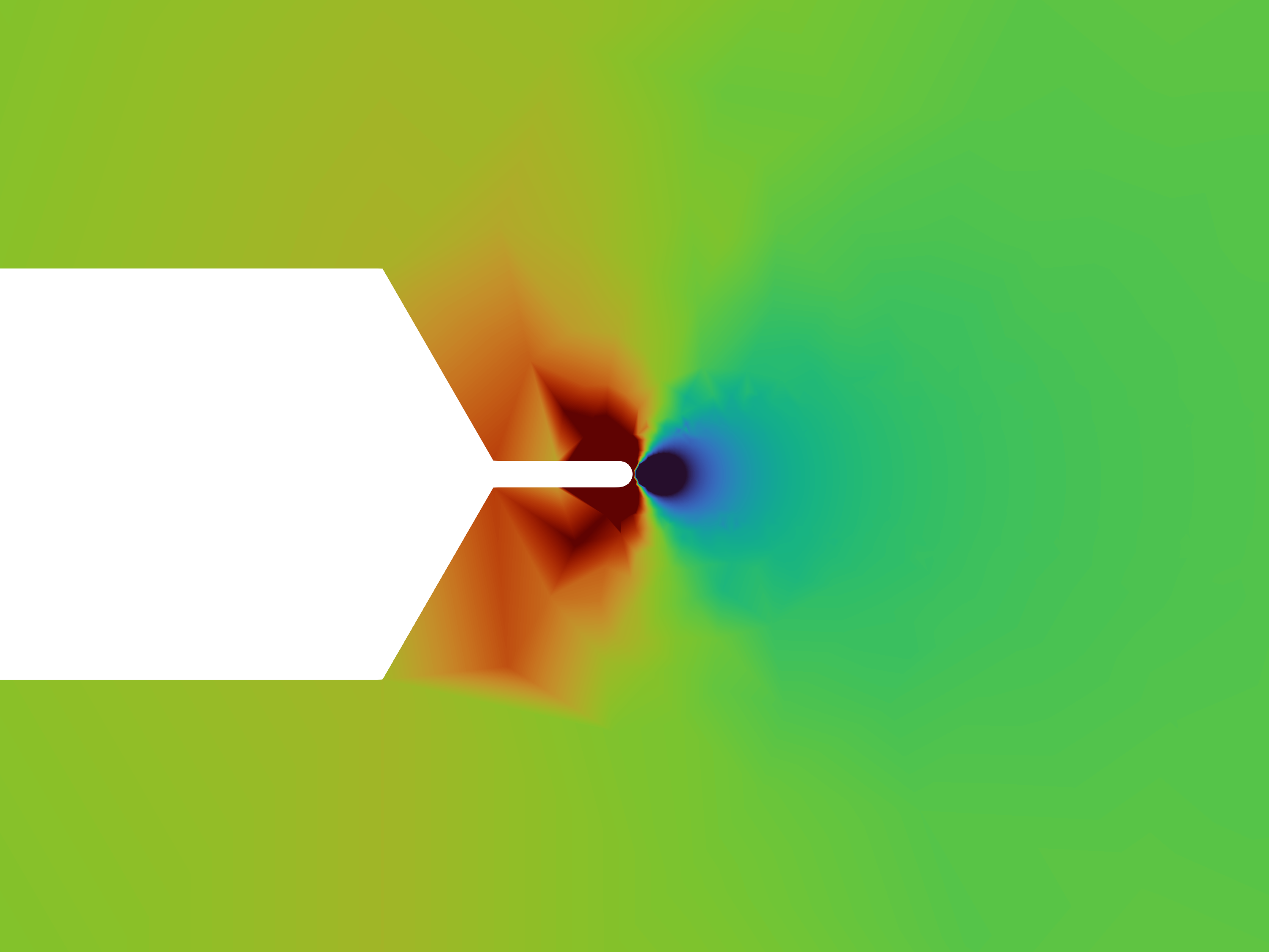}}; 
                \draw[black] (-0.25,0.381) rectangle (2.75,2.63);
                \draw[black] (-0.6,-0.1) -- (2.75,0.381);
                \draw[black] (-0.825,0.1) -- (-0.25,2.63);
                \node[align=right, anchor=south east] at (2.75,-2.63) {$N\!=\!84$, $\Delta N\!=\!3$\\$\min(\epsilon_{\bar{\alpha}})\!=\!-\!10^{-5}$\\$\max(\epsilon_{\bar{\alpha}})\!=\!10^{-5}$};
            \end{scope}
        \end{tikzpicture}
        \label{fig:prediction_error_field_II}
    }
    \subfloat[]{
        \begin{tikzpicture}[every node/.style={inner sep=0pt,outer sep=0pt}]
            \footnotesize
            \begin{scope}[scale=0.75]
                \node[anchor=center,fill=white] at (0,0) {\includegraphics[width=4.5cm]{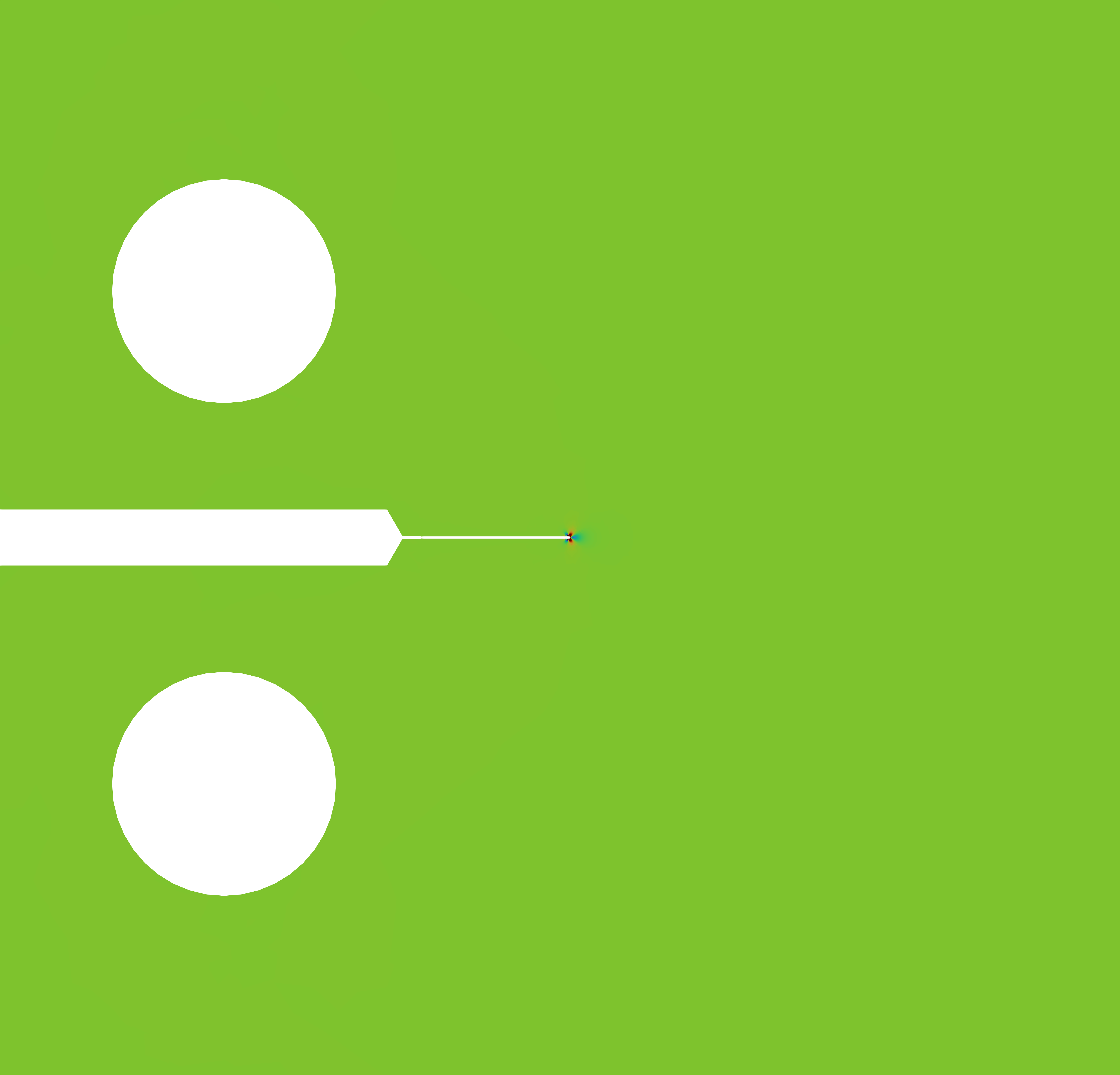}}; 
                \draw[black] (-0.575+0.35,-0.25) rectangle (0.075+0.35,0.25);
                \node[fill=white, anchor=north east] at (2.75,2.63) {\includegraphics[width=2.25cm]{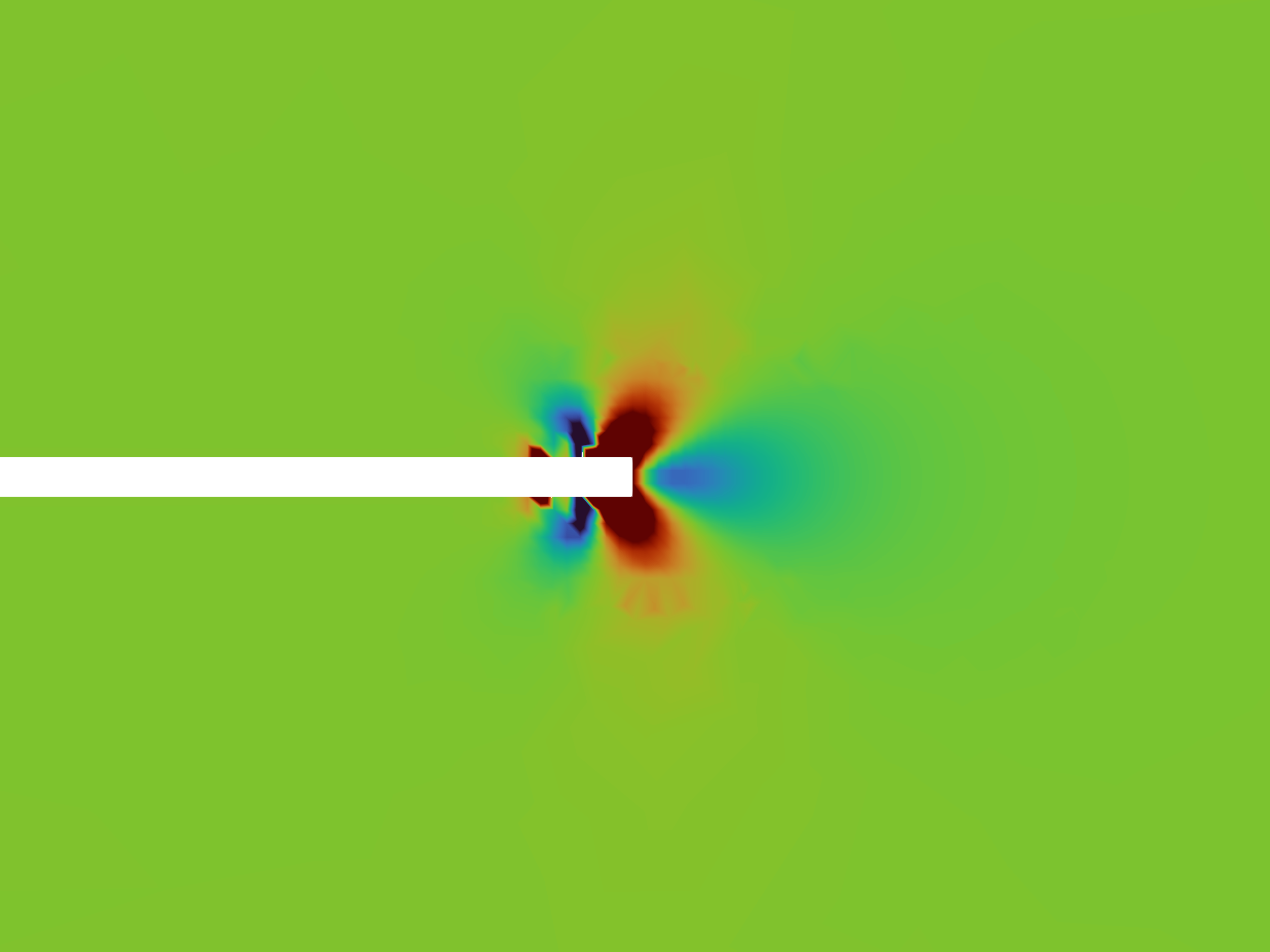}}; 
                \draw[black] (-0.25,0.381) rectangle (2.75,2.63);
                \draw[black] (0.075+0.35,0.25) -- (2.75,0.381);
                \draw[black] (-0.575+0.35,0.25) -- (-0.25,0.381);
                \node[align=right, anchor=south east] at (2.75,-2.63) {$N\!=\!1201$, $\Delta N\!=\!2$\\$\min(\epsilon_{\bar{\alpha}})\!=\!-\!10^{-2}$\\$\max(\epsilon_{\bar{\alpha}})\!=\!10^{-2}$};
            \end{scope}
        \end{tikzpicture}
        \label{fig:prediction_error_field_III}
    }
    \includegraphics{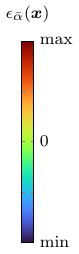}
    \caption{Approach to obtain an estimation of the prediction error (a) as well as representative prediction error field for a cycle jump in stage II (b) and stage III (c), where elements with $d\geq0.95$ are hidden to indicate the crack.}
    \label{fig:prediction_error_field}
\end{figure*}

Finally, we address the accuracy of the system state prediction, that is the extrapolation of $\bar{\alpha}$.
To estimate it, we compare the computed and extrapolated $\bar{\alpha} (\bs{x}, N)$ for the $N_{\text{s}}-1$ cycles following a cycle jump, i.e. for the cycles needed to perform the subsequent jump excluding the trial cycle, as illustrated in Fig. \ref{fig:prediction_error}.
For these cycles, we use  (\ref{eq:FD_extrapolation}) to compute an average error
\begin{equation}
    \begin{aligned}
        &\epsilon_{\bar{\alpha}} (\bs{x},N) = \frac{1}{N_{\text{s}}\!-\!1} \sum_{i=1}^{N_{\text{s}}-1}\\
        &\qquad\frac{\bar{\alpha}^\star (\bs{x}, N\!+\!\Delta N\!+\!i) - \bar{\alpha} (\bs{x}, N\!+\!\Delta N\!+\!i)}{\bar{\alpha} (\bs{x}, N\!+\!\Delta N\!+\!i)} \text{ .}
    \end{aligned}
\end{equation}
Figs. \ref{fig:prediction_error_field_II} and \ref{fig:prediction_error_field_III} show the contour plot of $\epsilon_{\bar{\alpha}} (\bs{x})$ for a representative cycle jump within stage II and stage III, respectively.
We adopt $P_{\max}=120$~N  since it exhibits the largest errors, but we obtain  similar results also at other load levels.
Clearly, the prediction error is localized around the notch in stage II or the crack tip  in stage III.
Crack propagation leads to a relocation of the zone with dominant fatigue effect, meaning that the FD-based prediction of $\bar{\alpha}$ leads to areas of under- and overestimation, as evident e.g. in Fig. \ref{fig:prediction_error_field_III}.

\begin{figure*}
    \centering
    \subfloat[]{
        \includegraphics{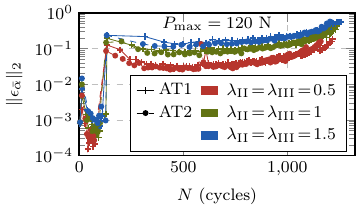}
        \label{fig:numerical_results_ct_error_L2_120}
    }
    \subfloat[]{
        \includegraphics{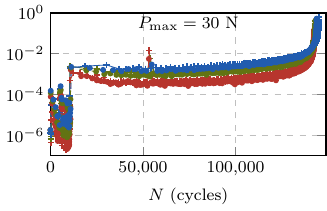}
        \label{fig:numerical_results_ct_error_L2_30}
    }
    \subfloat[]{
        \includegraphics{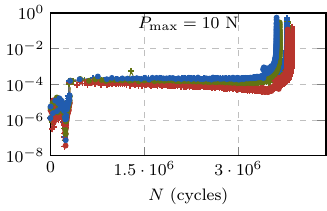}
        \label{fig:numerical_results_ct_error_L2_10}
    }
    \caption{$L2$-norm of the estimated prediction error throughout the fatigue life for the CT test at different representative load levels (a-c).}
    \label{fig:numerical_results_properties_error_L2}
\end{figure*}

Fig. \ref{fig:numerical_results_properties_error_L2} shows the evolution of the $L2$-norm of the estimated prediction error field $\norm{\epsilon_{\bar{\alpha}}}{2}$ at the three representative load levels.
Stage I and II cycle jumps exhibit smaller prediction errors than stage III cycle jumps, which we attribute to the relocation of the crack tip with crack propagation.
Only towards the end of the fatigue life close to $N_{\text{u}}$, we can observe a significant increase of the prediction error, which is due to the strongly non-linear behavior of the system in that stage.
However, the errors within each individual stage tend to be of the same order of magnitude, despite the very different cycle-jump sizes.
I.e., the cycle jumps in beginning of stage III at the lowest load level of more than $\Delta N \approx 10^5$~cycles lead to a similar prediction error as a cycle jump towards the end of this stage of $\Delta N = 2$~cycles.
This ultimately proves that a cycle-jump size based on the crack growth (or $\max \left( d (\bs{x}) \right)$) obtains stable prediction errors throughout the fatigue life, which was the original motivation for the cycle-jump criterion.
Further, we can observe that larger speedup factors $\lambda_{\text{II,III}}$ clearly yield a higher prediction error, in line with the previous explanations.
Finally, the AT1 and AT2 models yield very similar values for the prediction errors.
We remark that the estimation of the local state prediction error is obtained by post-processing the obtained data and is not used to modify the jump extension.
Although the checks on the accuracy of the prediction of the global quantities performed during the trial cycle (Section~\ref{sec:algorithmic_aspects} and \eqref{eq:correction_cycle_jump}) seem to be sufficient to guarantee a good accuracy, the introduction of a local error estimate can be a possibility to further improve the proposed ACJ.

\subsection{Perforated plate test}
\begin{figure}
    \centering
    \hspace{-0.5cm}
    \subfloat[]{
        \includegraphics{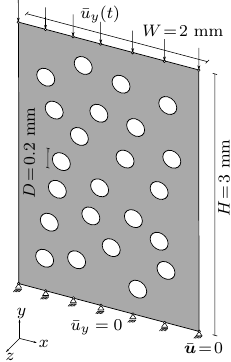}
        \label{fig:specimen_pwh}
    }
    \hspace{-0.7cm}
    \subfloat[]{
        \begin{tikzpicture}
                \node[anchor=center] at (0,0) {\hspace{0.5cm}\includegraphics[width=3.054cm]{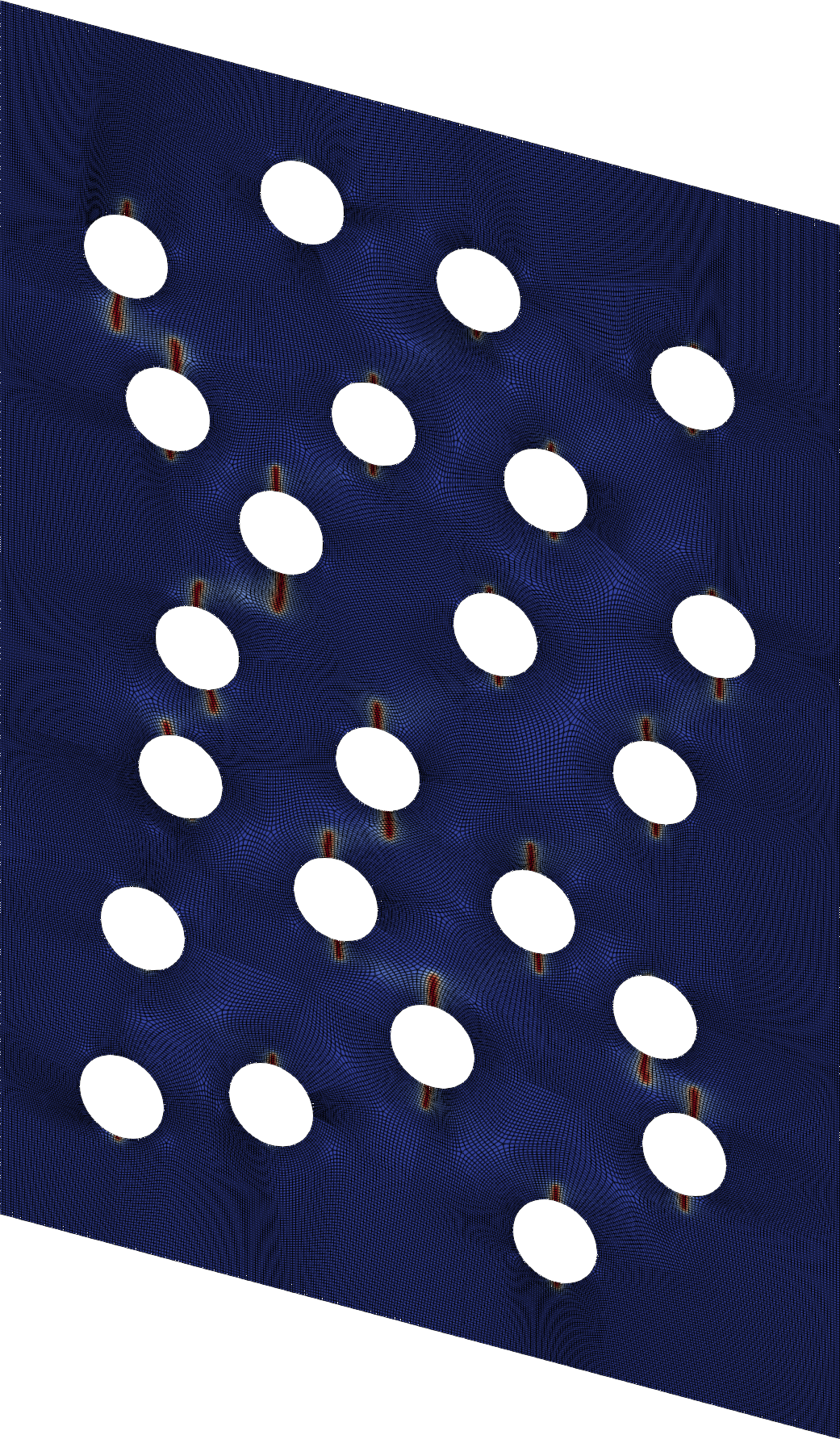}};
                \node[anchor=center] at (2.4,-0.2) {\includegraphics{colormap_pf.pdf}};
        \end{tikzpicture}\hspace{-0.5cm}
        \label{fig:phasefield_pwh}
    }
    \caption{Geometry and boundary conditions (a) as well as FE mesh showing the phase-field contour at an exemplary cycle during the stage of fatigue crack propagation (b) for the perforated plate test.}
    \label{fig:pwh_test}
\end{figure}
Next, a more complex geometry is tested to evaluate the proposed acceleration scheme for multiple cracks nucleating at different instants and propagating at different rates.
We use a setup similar to the one in \cite{nguyen_phase_2015}, illustrated in Fig. \ref{fig:specimen_pwh}. A plate with 23 randomly distributed holes under plane-strain conditions is subjected to a prescribed compressive cyclic displacement on the upper edge,
 $\bar{u}_y (t)$, with $\bar{u}_{y,\max} = 0$~mm and $\bar{u}_{y,\min} = -2.5\cdot10^{-4}, -5\cdot10^{-4}, -6.25\cdot10^{-4}, -7.5\cdot10^{-4}$~mm.
The adopted parameters are taken from \cite{carrara_framework_2020}, namely $E=12000$~MPa, $\nu=0.22$, $\mathcal{G}_{\text{c}} = 0.0014$~MPa~mm and $\ell = 0.018$~mm.
Also, we choose the AT2 model and the no-tension energy decomposition \cite{freddi_regularized_2010}, while we set $\bar{\alpha}_{\text{th}} = 0.00648$~N~mm$^{-2}$ and $p=2$ for the fatigue degradation function.
The mesh, shown in Fig. \ref{fig:phasefield_pwh}, features $121831$ bilinear quadrilateral elements and $123931$ nodes, leading to $\ell/h \approx 3$ in the whole domain.
For this test, we use six load steps per cycle, that is five during loading and one during unloading.
The challenge in this case is the presence of multiple cracks that nucleate at different cycles and propagate at different rates due to the randomness of the hole positions.
To show that the ACJ procedure is able to deal with this complexity without any modifications, again $\lambda_{\text{II}}=\lambda_{\text{III}}=0.5,1,1.5$ are used for the computations.
Since for this specific setup the definition of the failure point is not straightforward, all results presented here are obtained by means of a total computation wall time of 14 days.

\begin{figure*}
    \centering
    \subfloat[]{
        \includegraphics{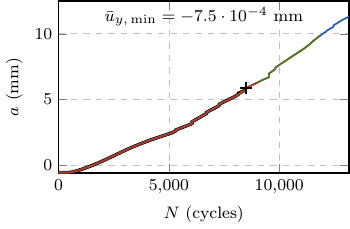}
        \label{fig:numerical_results_pwh_crackprop_7_5}
    }
    \subfloat[]{
        \includegraphics{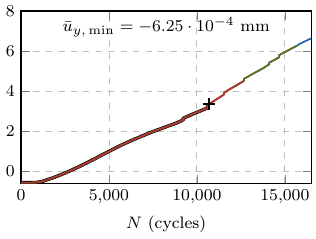}
        \label{fig:numerical_results_pwh_crackprop_6_25}
    }
    \subfloat[]{
        \includegraphics{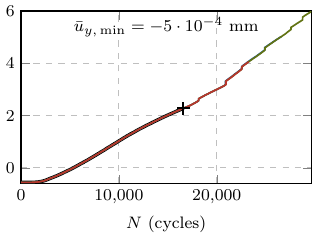}
        \label{fig:numerical_results_pwh_crackprop_5}
    }
    \newline
    \subfloat[]{
        \includegraphics{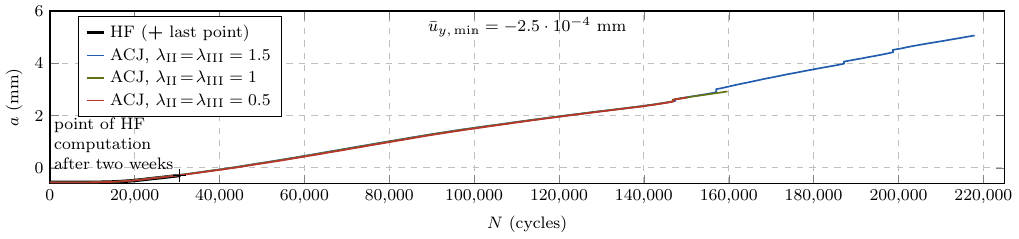}
        \label{fig:numerical_results_pwh_crackprop_2_5}
    }
    \caption{Comparison between the HF- and ACJ-computed total crack length as recorded by the smeared crack length approach for the perforated plate test with different maximum applied displacement magnitudes (a-d).}
    \label{fig:numerical_results_pwh_crackprop}
\end{figure*}

Fig. \ref{fig:numerical_results_pwh_crackprop} compares the total crack length, as measured by the smeared crack length concept (now setting the expected crack tip number to $k=23$\footnote{Note that the value used here for $k$ is a rough estimation of the expected number of crack tips, since in this case not all the cracks nucleate at the same instant and not each hole triggers two cracks. This approximation, however, does not influence significantly the results because of two reasons. First, we compute crack increments and, second, the contribution due to the crack extension is much larger than the tip one.}), for the four different load levels.
Evidently, the curves virtually coincide, while only the final predicted crack length is slightly overestimated by the proposed acceleration scheme, depending on the chosen speedup parameters.
The jumps in the smeared crack length indicate either the nucleation of new cracks or the merging of propagating cracks, which the ACJ scheme can resolve with virtually no deviation from the HF computations.
Despite this negligible difference, all of the ACJ computations reach a significantly higher cycle count in the 14 days of computation. This is especially evident at the lowest load amplitude with $\bar{u}_{y,\min} = -2.5\cdot10^{-4}$~mm (Fig. \ref{fig:numerical_results_pwh_crackprop_2_5}), where the ACJ computation with $\lambda_{\text{II,III}}=1.5$ reaches almost $220\,000$~cycles, in contrast to the HF computation which reaches around $30\,000$~cycles.

\begin{table*}
    \caption{Number of cycles associated to thecrack length attained by the HF computation within two weeks $N_{\text{co}}$, number of resolved cycles $N_{\text{r}}$ and CPU time comparing HF with ACJ computations for the perforated plate test.}
    \label{tab:PWH_compar}
    \centering
    \footnotesize
    \begin{tabularx}{\textwidth}{l|ll|lll|lll|lll}
        \toprule
        \multirow{3}{*}{$\bar{u}_{y,\min}$}
            &\multicolumn{2}{c}{HF}
            &\multicolumn{3}{c}{ACJ, $\lambda_{\text{II}}\!=\!\lambda_{\text{III}}\!=\!0.5$}
            &\multicolumn{3}{c}{ACJ, $\lambda_{\text{II}}\!=\!\lambda_{\text{III}}\!=\!1$}
            &\multicolumn{3}{c}{ACJ, $\lambda_{\text{II}}\!=\!\lambda_{\text{III}}\!=\!1.5$}\\
            &$N_{\text{co}}$ &time &$N_{\text{co}}$ &$N_{\text{r}}$ &time &$N_{\text{co}}$ &$N_{\text{r}}$ &time &$N_{\text{co}}$ &$N_{\text{r}}$ &time\\
            & &(h:min) & & &(h:min) & & &(h:min) & & &(h:min)\\
        \midrule
        $-7.5$	                              &8487       &330:42     &8488        &3754      &242:23   &8493       &2300        &168:57         &8475               &1652          &108:37     \\
        \hspace*{0.15cm}$\cdot10^{-4}$        &(100\%)    &(100\%)    &(+0.01\%)   &(44.23\%) &(73.3\%)   &(+0.07\%)  &(27.08\%)   &(51.09\%)      &(-0.14\%)          &(19.49\%)     &(32.84\%)  \\
        $-6.25$	                              &10679      &329:13     &10673       &2977      &193:37   &10672      &1708        &131:48         &10690              &1207          &106:39     \\
        \hspace*{0.15cm}$\cdot10^{-4}$        &(100\%)    &(100\%)    &(-0.06\%)   &(27.89\%) &(58.82\%)   &(-0.07\%)  &(16.0\%)    &(40.04\%)      &(+0.10\%)          &(11.29\%)     &(32.4\%)   \\
        $-5$	                              &16583      &326:51     &16582       &2592      &136:52   &16610      &1378        &63:20          &16574              &937           &50:45      \\
        \hspace*{0.15cm}$\cdot10^{-4}$        &(100\%)    &(100\%)    &(-0.01\%)   &(15.63\%) &(41.88\%)   &(-0.16\%)  &(8.3\%)     &(19.38\%)      &(-0.05\%)          &(5.65\%)      &(15.53\%)  \\
        $-2.5$	                              &30603      &328:27     &30001       &568       &9:32   &29598      &288         &3:38           &29323              &204           &2:46       \\
        \hspace*{0.15cm}$\cdot10^{-4}$        &(100\%)    &(100\%)    &(-1.97\%)   &(1.89\%)  &(+2.9\%)  &(-3.28\%)  &(0.97\%)    &(1.11\%)       &(-4.18\%)          &(0.7\%)       &(0.84\%)   \\
        \bottomrule
    \end{tabularx}
\end{table*}

In this case, to compare the accelerated and HF results we consider the cycle $N_{\text{co}}$ associated to the crack length $a_{\text{co}}$ which the HF computations could obtain within two weeks of computational time.
Tab. \ref{tab:PWH_compar} gives a comparison in terms of $N_{\text{co}}$, number of resolved cycles $N_{\text{r}}$ and CPU time for the different speedup factors.
All the parameter combinations achieve an error of $N_{\text{co}}$ below around $3$~\%, while only computing a small fraction of the total number of cycles.
At the highest load amplitude, the system evolution is so quick that almost no cycle can be jumped, while for the lowest load level we notice a high speedup.
For $\bar{u}_{y,\min} = -2.5\cdot10^{-4}$~mm, the ACJ with $\lambda_{\text{II,III}}=1.5$ reaches the reference point of the HF computations within less than three hours of computational time (less than one percent of the HF time), while resolving only $204$~cycles ($0.7$~\%).
Continuing after this reference point, it reaches $217\,864$~cycles by resolving only $2033$~cycles ($0.93$\%) within around $14$~days of computational time.

\begin{figure*}
    \centering
    \subfloat[]{
        \includegraphics{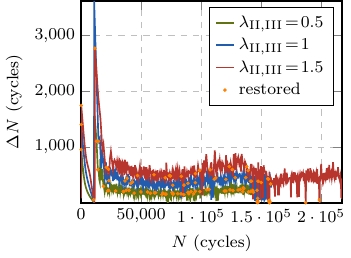}
        \label{fig:numerical_results_pwh_cyclejump_DN}
    }
    \subfloat[]{
        \includegraphics{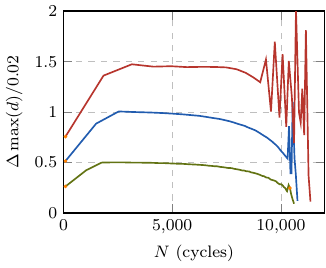}
        \label{fig:numerical_results_pwh_cyclejump_Dmaxd}
    }
    \subfloat[]{
        \includegraphics{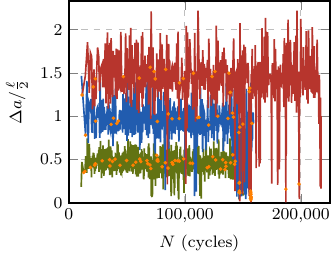}
        \label{fig:numerical_results_pwh_cyclejump_Da}
    }
    \caption{Comparison of the size of the performed cycle jumps $\Delta N$ (a), the obtained increments $\Delta \max(d)$ in stage II (b), and the obtained total crack length advancements $\Delta a$ (c) for the perforated plate test with $\bar{u}_{y,\min} = -2.5\cdot10^{-4}$~mm. The data points are connected by lines for better visibility.}
    \label{fig:pwh_cyclejump}
\end{figure*}

Next, Fig. \ref{fig:pwh_cyclejump} shows the performed cycle jumps throughout the fatigue life $\Delta N (N)$ as well as the increments of $\Lambda$ during the jumps in stage II and stage III at $\bar{u}_{y,\min} = -2.5\cdot10^{-4}$~mm.
In comparison to samples with a single crack tip, stronger oscillations are observed, however the target increments $\Delta \max (d)$ and $\Delta a$ are still around the target value $\Delta \bar{\Lambda}$.
As indicated with the orange points, the rejection of a cycle jump with too large $\Delta \Lambda$ or not converging occurs slightly more often than for the single-crack case, but is still very limited.

\begin{figure}
    \centering
    \includegraphics{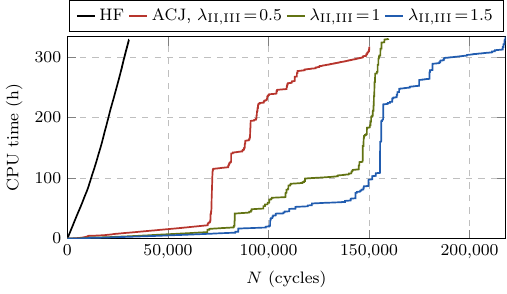}
    \caption{Comparison between the computational time as a function of the cycle count for the perforated plate test with $\bar{u}_{y,\min} = -2.5\cdot10^{-4}$~mm, for HF and ACJ computations with different speedup parameters.}
    \label{fig:numerical_results_pwh_cputime}
\end{figure}

Finally, Fig. \ref{fig:numerical_results_pwh_cputime} shows the CPU time as a function of the cycle count $N$ for the lowest load amplitude with $\bar{u}_{y,\max} = -2.5\cdot10^{-4}$~mm.
The different slope of the curves for the HF and the ACJ computations with different speedup parameters clearly demonstrates the obtained speedup.
While the HF computations show an almost linear correlation between computational time and cycle count, the ACJ results comprise regions where the computations appear to be much slower (at around $150\,000$~cycles). This is due to a strongly nonlinear and hence barely predictable evolution of the system during the nucleation of new cracks or the merging of existing cracks (Fig. \ref{fig:numerical_results_pwh_crackprop_2_5}). There, the scheme provides reduced jump extensions (Fig.~\ref{fig:numerical_results_pwh_cyclejump_DN}) and it needs to restore cycles more frequently (Fig. \ref{fig:numerical_results_pwh_cyclejump_Da}).
This demonstrates the robustness of the proposed acceleration scheme and that it is capable of dealing with complex behaviors thanks to the definition of the global variables constraining the jump.

\section{Conclusions}
\label{sec:conclusions}
In this work, we propose an adaptive cycle-jump scheme to address the prohibitive computational cost of phase-field fatigue computations associated with predictions in the high cycle fatigue regime.
We introduce a cycle-jump criterion which allows to determine when a jump is feasible and how many cycles can be skipped based on a target increment of a global representative variable during the skipped cycles.
For the definition of this global variable, the fatigue life is divided into (I) an initial stage before the fatigue effect is triggered, (II) the stage of fatigue crack nucleation, and (III) the crack propagation stage.
For the third stage, to overcome issues of conventional crack tip tracking algorithms, we introduce the smeared crack length concept which is based on a comparison of the integral of the numerically obtained phase-field solution with the integral of the optimal phase-field profile.
Based on representative numerical tests, the following conclusions can be drawn:
\begin{itemize}
    \item in contrast to existing acceleration techniques, the proposed approach shows a consistent behavior across different fatigue regimes as well as different loading and test scenarios;
    \item the number of cycles to be resolved depends only on a few parameters with clear meaning and can  be estimated with good accuracy;
    \item the scheme automatically reverts back to a HF resolution of the cycles once fast system evolution is detected, thereby preventing loss of accuracy during critical stages of the fatigue life;
    \item errors below $3$~\% in terms of total fatigue life and a very good agreement of crack growth rate predictions can be achieved with a speedup factor of up to more than four orders of magnitude, which can directly be influenced by optional user parameters;
    \item thanks to the smeared crack length approach, the scheme is able to deal with complex phase-field fatigue computations such as with specimens including multiple cracks.
\end{itemize}
Ultimately, the proposed scheme enables phase-field fatigue computations in the HCF and VHCF regimes which were simply not possible without acceleration.

\backmatter

\bmhead{Supplementary information}

\bmhead{Acknowledgements}
The help of Francesco Vicentini with the computation of the phase-field profiles during localization is gratefully acknowledged.
We acknowledge funding from the Swiss National Science Foundation through Grant No. 200021-219407 `Phase-field modeling of fracture and fatigue: from rigorous theory to fast predictive simulations'.
Jonas Heinzmann acknowledges funding of the German Academic Exchange Service (DAAD) for a scholarship in the PROMOS programme.

\bmhead{Code Availability}
The implementation of the adopted phase-field fatigue model, the cyclic solver as well as the proposed cycle-jump scheme are available in the open-source library GRIPHFiTH \cite{carrara_griphfith_2023} and can be accessed at \url{https://gitlab.ethz.ch/compmech/GRIPHFiTH}.

\begin{appendices}

\section{Correction and validation of the smeared crack length approach}
In the following, we investigate and correct the influence of a non-optimal phase-field profile on the smeared crack length approach; then, we use a representative numerical test to validate it.

\subsection{Correction of the smeared crack length}\label{sec:corrections_of_the_smeared_crack_length}
As explained in Section \ref{sec:smeared_crack_length_approach}, the assumption of an optimal phase-field profile may not necessarily hold in the numerical context, leading to a wrong estimation of the crack length.
This is due to the following reasons:
\begin{itemize}
    \item the irreversibility condition of the phase field $\dot{d}~\geq~0$;
    \item the phase field around the crack tip does not correspond to the revolved optimal half-profile, as also mentioned by \citeauthor{freddi_fracture_2019} in \cite{freddi_fracture_2019}. The effect of the crack tip on the phase-field regularization is systematically studied in \cite{pascale_systematic_2023}, and its influence is shown to be especially large for smaller cracks;
    \item the discretization error;
    \item the phase-field fatigue model, by locally degrading the fracture toughness of the material, leads to a heterogeneous $\mathcal{G}_{\text{c}}$, whose effect on the phase-field profile is studied in \cite{vicentini_phase-field_2023}.
\end{itemize}
In the following, we introduce corrections for the first three systematic and predictable errors.

\subsubsection{Error due to damage irreversibility}
\begin{figure}[t]
    \centering
    \subfloat[]{
        \includegraphics{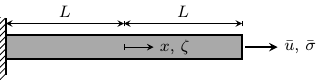}
        \label{fig:1D_bar}
    }\\[-0.5em]
    \subfloat[]{
        \includegraphics{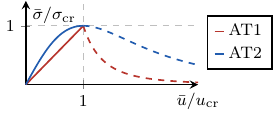}
        \label{fig:solution_homogeneous}
    }
    \subfloat[]{
        \includegraphics{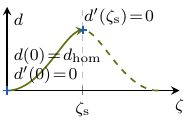}
        \label{fig:solution_localized}
    }
    \caption{Methodology to obtain the phase-field profiles during localization for a 1D bar (a) with its homogeneous solution (b) and constraints for the semi-analytical solution (c).}
    \label{fig:irrev_profile_method}
\end{figure}

First, we study the influence of the irreversibility condition $\dot{d}\geq 0$.
We consider a 1D bar of length $2L\gg 2\ell$, clamped at one end and subjected to a prescribed displacement $\bar{u}$ at the other end, see Fig. \ref{fig:1D_bar}.
The behavior of the bar is thoroughly discussed by \citeauthor{pham_approche_2010} in \cite{pham_approche_2010,pham_gradient_2011,pham_onset_2013}; here we only briefly outline some main results which are useful for the purpose of our correction. First, starting from an undamaged condition and loading the bar under displacement control, we obtain a solution with homogeneous damage, reported in terms of normalized stress-displacement response in Fig. \ref{fig:solution_homogeneous}. When the peak stress is reached, the phase field attains the value \cite{pham_gradient_2011}
\begin{equation}
    d_{\text{hom}} =
    \begin{cases}
        0    & \text{AT1} \\
        0.25 & \text{AT2} \\
    \end{cases}
\end{equation}
after which the homogeneous solution becomes unstable (assuming a sufficiently long bar, $L \gg \ell$).
At this point, damage localization occurs at an arbitrary location in the bar.
During progressive localization, the maximum phase-field value increases monotonically and a snap-back occurs, so that following the behavior requires a switch to (decreasing) stress control with $\bar{\sigma}$.
As follows, we compute the intermediate localization profiles during this stage, starting with the homogeneous solution $d=d_{\text{hom}}$ valid at the stress peak, and define their envelope as the `irreversible' phase-field profile. This allows us to then evaluate the discrepancy between this profile and the optimal profile, which can be obtained analytically by ignoring the irreversibility condition (see e.g. \cite{gerasimov_penalization_2019} for the detailed derivation). 

To do so, we adopt a semi-analytical approach similar to those in \cite{pham_construction_2009,vicentini_phase-field_2023}. We start by writing the damage criterion (\ref{eq:pf_KKT}), here considered as an equality since the phase field is evolving, i.e. $\dot{d} > 0$.
Due to the stress control required during the localization stage, the damage criterion is expressed in terms of stress by introducing the compliance $S(d) := 1/E(d)$, with $E(d)=g(d)E_0$, which leads to
\begin{equation}
    \frac{1}{2} S^\prime(d) \bar{\sigma}^2 = \frac{\mathcal{G}_{\text{c}}}{c_w} \left( \frac{w^\prime (d)}{\ell} - 2 \ell d^{\prime\prime} \right) \text{ .}
\end{equation}
Introducing the non-dimensional coordinate $\zeta:=x/\ell$ and the non-dimensional stress $\varrho := \bar{\sigma} / \sigma_{\text{cr}}$, with
\begin{equation}
    \sigma_{\text{cr}} =
    \begin{cases}
        \hphantom{\frac{9}{16}} \sqrt{\frac{3 E \mathcal{G}_{\text{c}}}{8 \ell}} & \text{AT1} \\
        \frac{3}{16} \sqrt{\frac{3 E \mathcal{G}_{\text{c}}}{\ell}}            & \text{AT2} \\
    \end{cases} \text{,}
\end{equation}
yields the non-dimensional damage criterion
\begin{equation}
    \begin{cases}
        \frac{1}{(1-d)^3} \varrho^2 = 1 - 2 d^{\prime\prime}                             & \text{AT1} \\
        \frac{27}{128(1-d)^3} \varrho^2 = 2d - 2 d^{\prime\prime} & \text{AT2} \\
    \end{cases}
\end{equation}
which we solve numerically for half of the domain with initial conditions
\begin{equation}
    d (\zeta = 0) = d_{\text{hom}} \quad\text{and}\quad d^\prime (\zeta = 0) = 0 \text{ .}
\end{equation}
Here we are assuming that outside of the localization zone, the phase field takes the value of the homogeneous solution at peak stress $d_{\text{hom}}$, due to the irreversibility of the damage accumulated during the homogeneous stage.
This is in contrast to the reversible case considered in \cite{pham_gradient_2011}, where the phase field is assumed to vanish outside the localization zone.
For the AT2 model, the initial condition of a vanishing first derivative is relaxed by setting $d^\prime(0) = 10^{-12}$ to avoid numerical difficulties due to the infinite support.
Since we do not know the width of the localization profile, we need one additional condition; denoting the half-width as $\zeta_s$, we impose $d^\prime(\zeta_s)=0$ (see Fig. \ref{fig:solution_localized}).

\begin{figure*}
    \centering
    \subfloat[]{
        \includegraphics{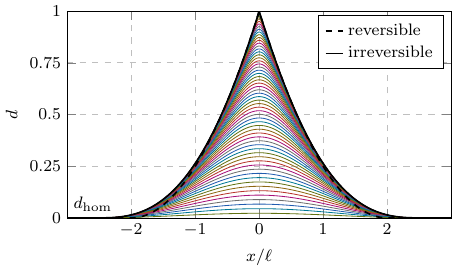}
        \label{fig:phasefield_profile_irreversible_AT1_profiles}
    }
    \subfloat[]{
        \includegraphics{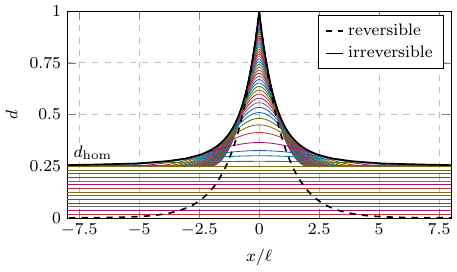}
        \label{fig:phasefield_profile_irreversible_AT2_profiles}
    }
    \newline
    \subfloat[]{
        \includegraphics{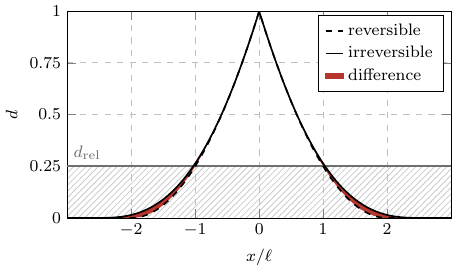}
        \label{fig:phasefield_profile_irreversible_AT1_diff}
    }
    \subfloat[]{
        \includegraphics{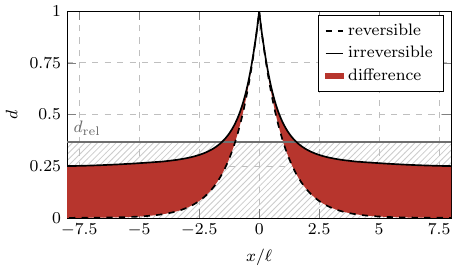}
        \label{fig:phasefield_profile_irreversible_AT2_diff}
    }
    \caption{Irreversible phase-field profiles and deviation from optimal profile for the AT1 (a, c) and AT2 (b, d) model, respectively.}
    \label{fig:phasefield_profile_irreversible}
\end{figure*}
The irreversible phase-field profiles obtained with this procedure and centered in the middle of the bar are depicted in Fig. \ref{fig:phasefield_profile_irreversible}.
As evident for the AT1 model in Fig. \ref{fig:phasefield_profile_irreversible_AT1_profiles}, the support of the irreversible profile is slightly wider than that of the optimal profile, and the difference in area is $+4.2$~\%, filled in red in Fig. \ref{fig:phasefield_profile_irreversible_AT1_diff}. Only for values of $d \gtrapprox 0.5$ the two profiles nearly coincide.
A similar but much stronger effect can be observed for the AT2 model, see Fig. \ref{fig:phasefield_profile_irreversible_AT2_profiles}, where the large homogeneous solution causes major deviations from the optimal phase-field profile. In this case, it is only for values of $d \gtrapprox 0.75$ that the two profiles nearly coincide.
Clearly, this behavior depends on the homogeneous solution; while this is known for the 1D bar, it is no longer valid for a general 2D or 3D case.
Further, since the localized solutions occur during a snap-back, they cannot be captured by a numerical solution scheme without a stress-controlled or an arc-length controlled solver.
However, they are more likely to occur in the fatigue context due to the continuous decrease of the fracture toughness.

Due to the dependency of the areal difference on the homogeneous solution which cannot be determined \textit{a priori} in a general case, instead of introducing a factor to correct the smeared crack length approach, we adopt the approach of ignoring the portion of the phase-field profile below a threshold $d_{\text{rel}}$, such that the effect of irreversibility in the remaining profile is negligible.
As depicted in in Figs. \ref{fig:phasefield_profile_irreversible_AT1_diff} and \ref{fig:phasefield_profile_irreversible_AT2_diff}, we choose $d_{\text{rel}}$ as the level of damage at $x=\pm\ell$ for the optimal profile
\begin{equation}
    d_{\text{rel}} =
    \begin{cases}
        0.25                         & \text{AT1} \\
        \text{exp}(-1) \approx 0.368 & \text{AT2} \\
    \end{cases} \text{ .}
    \label{eq:d_rel}
\end{equation}
This choice represents a compromise between a small influence of the irreversibility on the compared profiles and the robustness of the approach, since the higher the threshold value, the less points of the numerical solution contribute to the numerical phase-field integral $D_{\text{num}}$.
Accordingly, the modified definition of the smeared crack length is
\begin{equation}
    a =
    \begin{cases}
        \displaystyle\frac{D_{\text{num}} (d) - k \pi \frac{11}{48} \ell^2}{\frac{7}{6} \ell}                 & \forall d \geq d_{\text{rel}} \qquad \text{AT1} \\
        \displaystyle\frac{D_{\text{num}} (d) - k \pi \ell^2 (1-2\text{exp}(-1))}{2 \ell (1- \text{exp}(-1))} & \forall d \geq d_{\text{rel}} \qquad \text{AT2} \\
    \end{cases} \text{ .}
\end{equation}

\subsubsection{Crack tip and discretization error}
\begin{figure}
    \centering
    \includegraphics{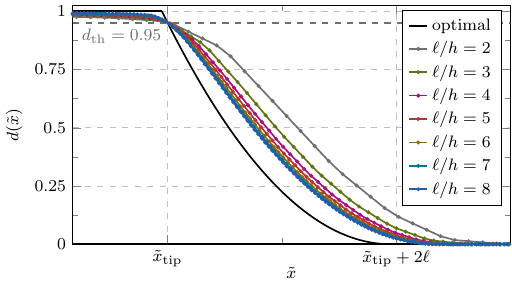}
    \caption{Study of the crack tip and discretization error with the interpolated phase field along the crack path $\tilde{x}$.}
    \label{fig:regularization_discretization_error}
\end{figure}
Next to the irreversibility, the crack tip and discretization errors cause deviations from the optimal phase-field profile \cite{freddi_fracture_2019,pascale_systematic_2023}, which are illustrated in Fig. \ref{fig:regularization_discretization_error}.
Evidently, the profile around the crack tip is not only different from the optimal profile, but it also depends on the FE mesh (discretization error).
These errors cannot be corrected exactly due to the inherent numerical approximation, but their systematic influence can be minimized by introducing empirical correction factors.
We use one correction factor for the crack tip contribution $c_{\text{tip}}$ and one for the contribution of the crack length $c_{\text{ext}}$,
\begin{equation}
    a = \frac{D_{\text{num}} (d) - k c_{\text{tip}} D_{\text{tip}}(\ell)}{c_{\text{ext}} D_{\text{ext}} (\ell)} \text{ .}
    \label{eq:aell_correction}
\end{equation}
\begin{figure}
    \centering
    \includegraphics{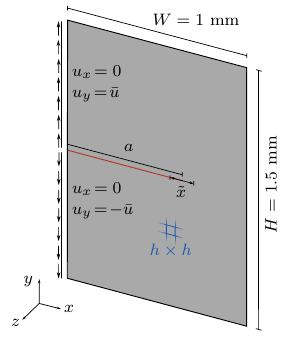}
    \caption{Geometry and boundary conditions for the specimen to determine the correction factors.}
    \label{fig:specimen_pulltest}
\end{figure}
To determine these correction factors empirically, we use the specimen in Fig. \ref{fig:specimen_pulltest}.
On the left boundary, the displacement is fixed in the $x$-direction, while the upper and lower halves of the left edge are pulled apart under displacement control to obtain the nucleation and propagation of a straight crack in the middle of the specimen.
We adopt $E=210$~MPa, $\nu=0.3$, $\mathcal{G}_{\text{c}} = 0.0027$~MPa~mm), no energy decomposition and plane-strain conditions.
Furthermore, we use the penalty method \cite{gerasimov_penalization_2019} with $\mathtt{TOL}_{\text{ir}}=10^{-6}$ to enforce irreversibility of the phase field, and compute the smeared crack length using the $L1$-norm approximation (\ref{eq:D_num_L1_approx}) due to the uniform discretization of the phase-field support.
To provide a comprehensive study of the crack tip and discretization error, all combinations of $\ell = 0.01, 0.02, 0.03, 0.04, 0.05, 0.06$~mm and the element sizes $\ell/h = 2, 3, 4, 5, 6, 7, 8$ are investigated.
Although the outer dimensions are chosen large enough with respect to $\ell$, we only consider the phase-field solution in the subdomain with $x>W/5$ and up to a crack length of $a=W/2$ to avoid boundary effects.
We use the interpolated crack tip method outlined in Fig. \ref{fig:cracktip_interpolation} to calibrate the smeared crack length approach.
In particular, the correction factor for the crack tip contribution $c_{\text{tip}}$ is computed by comparing the point at which the first full crack development is obtained, while the correction factor for the crack extent $c_{\text{ext}}$ is determined such that the crack growth rates coincide in a best-fit sense.

\begin{table}
    \caption{Correction factors for the crack tip and discretization error, obtained with bilinear quadrilateral elements, $d_{\text{th}}=0.95$ and considering irreversibility by means of (\ref{eq:d_rel}).}
    \label{tab:correction_factors}
    \centering
    \footnotesize
    \begin{tabularx}{\linewidth}{lXXX}
        \toprule
        regularization       & $\ell/h$ & $c_{\text{tip}}$ & $c_{\text{ext}}$\\
        \midrule
        \multirow{7}{*}{AT1}    & $2$      & $3.171$              & $1.582$\\
                                & $3$      & $2.244$              & $1.390$\\
                                & $4$      & $1.846$              & $1.302$\\
                                & $5$      & $1.680$              & $1.249$\\
                                & $6$      & $1.539$              & $1.211$\\
                                & $7$      & $1.468$              & $1.187$\\
                                & $8$      & $1.416$              & $1.171$\\
        \midrule
        \multirow{7}{*}{AT2}    & $2$      & $3.931$              & $1.604$\\
                                & $3$      & $2.875$              & $1.411$\\
                                & $4$      & $2.410$              & $1.323$\\
                                & $5$      & $2.136$              & $1.271$\\
                                & $6$      & $2.014$              & $1.228$\\
                                & $7$      & $1.893$              & $1.204$\\
                                & $8$      & $1.812$              & $1.185$\\
        \bottomrule
    \end{tabularx}
\end{table}

\begin{figure*}
    \centering
    \subfloat[]{
        \includegraphics{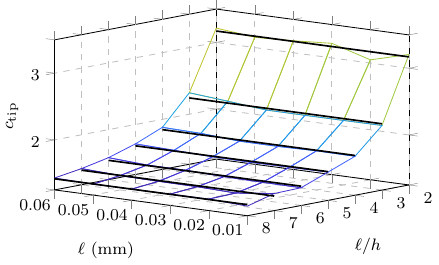}
        \label{fig:aell_correction_AT1_correction_tip}
    }\subfloat[]{
        \includegraphics{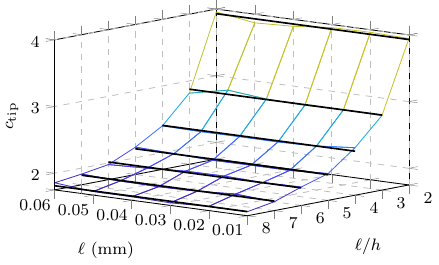}
        \label{fig:aell_correction_AT2_correction_tip}
    }
    \newline
    \subfloat[]{
        \includegraphics{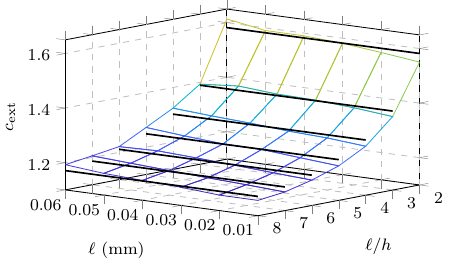}
        \label{fig:aell_correction_AT1_correction_ext}
    }
    \subfloat[]{
        \includegraphics{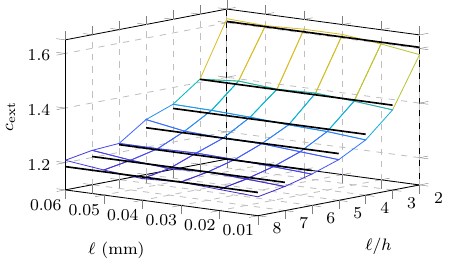}
        \label{fig:aell_correction_AT2_correction_ext}
    }
    \caption{Correction factors of tip and length contributions for the smeared crack length approach as a function of the length scale $\ell$ and the $\ell/h$ ratio with the AT1 (a, c) and AT2 model (b, d), respectively. The data points are connected with lines for better visibility.}
    \label{fig:aell_correction}
\end{figure*}

The obtained correction factors are visualized in dependency of the regularization parameter $\ell$ and the discretization parameter $\ell/h$ in Fig. \ref{fig:aell_correction}.
We can observe that the correction factors are virtually independent of $\ell$, while all of them show the same trend in the direction of $\ell/h$, converging towards one value with increasing $\ell/h$.
For the various $\ell/h$, we take the correction factors $c_{\text{tip}}$ and $c_{\text{ext}}$ as average of the values for different length scale parameters $\ell$, as indicated with the black lines in Fig. \ref{fig:aell_correction}.
They are listed in Tab. \ref{tab:correction_factors}, meaning that the correction of the smeared crack length must be set according to the chosen discretization for a most accurate measurement of the smeared crack length approach.
These values were obtained already accounting for the irreversibility, using bilinear quadrilateral elements. We remark that the correction factors might depend on the particular FE implementation and chosen tolerances.
Also, other element types and orders may require a different correction factor.

\subsection{Validation of the smeared crack length approach}\label{sec:validation_of_the_smeared_crack_length}
Finally, we validate the concept of the smeared crack length by comparing it to the discrete and interpolated crack tip tracking algorithms outlined in Fig. \ref{fig:cracktip_interpolation}.
In particular, we use the CT test setup of Section \ref{sec:behavior_of_the_proposed_approach} and depicted in Fig. \ref{fig:ct_test} for a fatigue computation at the load level $P_{\max} = 110$~N.
Further, we compare the smeared crack length computed by numerically integrating the phase-field solution to its approximation using the $L1$-norm of the phase-field solution.
As before, the interpolated crack tip method (Fig. \ref{fig:cracktip_interpolation}) is considered to be the most accurate in cases where the crack path is straight and known \textit{a priori}, such as for the test at hand.

\begin{figure*}
    \centering
    \includegraphics{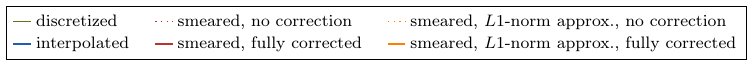}\\
    \subfloat[]{
        \includegraphics{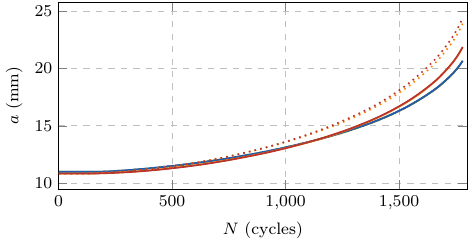}
        \label{fig:aell_validation_AT1_a}
    }
    \hfill
    \subfloat[]{
        \includegraphics{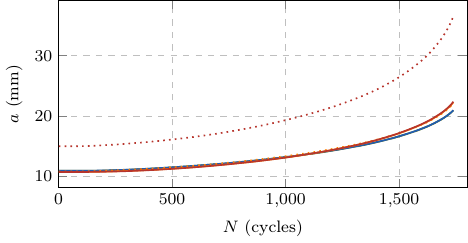}
        \label{fig:aell_validation_AT2_a}
    }
    \newline
    \subfloat[]{
        \includegraphics{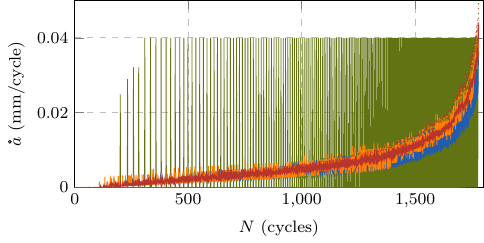}
        \label{fig:aell_validation_AT1_da}
    }
    \hfill
    \subfloat[]{
        \includegraphics{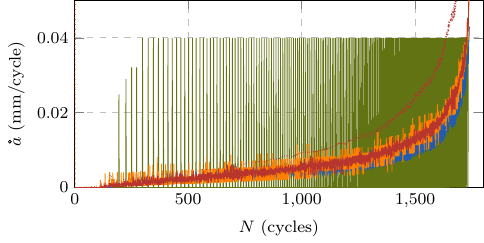}
        \label{fig:aell_validation_AT2_da}
    }
    \newline
    \subfloat[]{
        \includegraphics{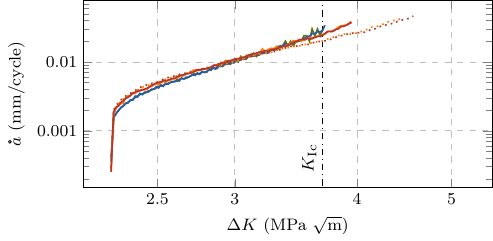}
        \label{fig:aell_validation_AT1_cgr}
    }
    \subfloat[]{
        \includegraphics{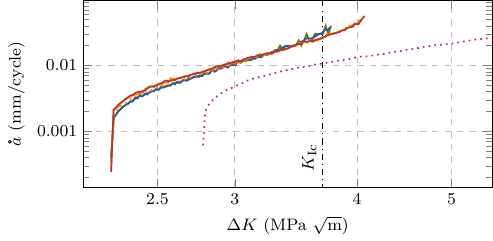}
        \label{fig:aell_validation_AT2_cgr}
    }
    \caption{Validation of the smeared crack approach with a CT test in terms of crack length (a, b), crack growth (c, d) and crack growth rate curve (e, f), as obtained with the different crack tip tracking algorithms for the AT1 (a,c,e) and the AT2 model (b,d,f), respectively.}
    \label{fig:aell_validation}
\end{figure*}

Comparing the measured crack lengths of the three different approaches in Figs. \ref{fig:aell_validation_AT1_a} and \ref{fig:aell_validation_AT2_a}, we can observe that the discrete and interpolated crack lengths virtually coincide.
The smeared crack length in its uncorrected form shows a significant deviation especially towards the end of the fatigue life, which can be attributed to the deviation due to the irreversibility as well as the regularization and crack tip error accumulating over time.
For the AT2 model additionally, a significant deviation can be observed already from the beginning, which is due to the homogeneous solution acting as an artificial initial crack length.
Accounting for all these factors, the smeared crack length measurements in their corrected form show a good agreement with the other two crack tip tracking algorithms.
The remaining deviations of the corrected smeared crack length are attributed to the facts that the correction factors were obtained with homogeneous distribution of the fracture toughness and with the penalty method \cite{gerasimov_penalization_2019}, whereas here the fatigue effects lead to a heterogeneous $f(\bar{\alpha})\mathcal{G}_{\text{c}}$ and the history-variable approach \cite{miehe_phase_2010} is used for irreversibility.
The approximation of the smeared crack length approach with the $L1$-norm shows a good agreement with the smeared crack length obtained with numerical integration proving its validity, which is due to the almost perfectly uniform discretization of the phase-field support in this specific test.

Comparing the measured crack growth rates $\dn{a}$ in Figs. \ref{fig:aell_validation_AT1_da} and \ref{fig:aell_validation_AT2_da}, the differences of the compared approaches become apparent:
the discrete crack length monitors exclusively a crack growth rate corresponding to the element size (here mostly $\mathrm{d}a = 0.04$~mm), while in most cycles, it records $\mathrm{d}a = 0$~mm.
In contrast, the interpolated crack tip tracking approach yields a more precise measurement of the crack growth rate since the crack is straight and known \textit{a priori}, although the measurement is subject to oscillations from cycle to cycle.
The smeared crack approach matches the interpolated crack growth rate very well, while the uncorrected smeared crack length overestimates the crack growth rate especially towards the end of the fatigue life.
The smeared crack length gives overall the most stable trend, hence making it suitable for the novel acceleration scheme which relies on a robust crack length measurement in stage III, while its approximation with the $L1$-norm is subject to slightly more instabilities.
This is why we use the smeared crack length as computed with numerical integration for the proposed cycle-jump scheme, since the LLSQ fit and thus the algorithm performs better if the underlying crack growth rate evaluation is more robust.

To finally showcase the capability of the smeared crack length concept, the crack growth rate curves are shown in Figs. \ref{fig:aell_validation_AT1_cgr} and \ref{fig:aell_validation_AT2_cgr} for the different crack tip tracking methods.
While the predictions from the discrete and interpolated crack tip show good agreement, with the former showing slightly more oscillations, the smeared crack length approach matches both conventional approaches with only slight deviations towards the final stage of unstable crack propagation.
Again, the uncorrected smeared crack length shows significant deviations especially for the AT2 dissipation function, highlighting the importance of the correction factors.
Overall, these results emphasize that the smeared crack length concept in its corrected form can substitute a conventional crack tip tracking algorithm.

\end{appendices}

\bibliography{references}

\end{document}